\documentclass[pra,twocolumn,showpacs,preprintnumbers,amsmath,amssymb,tightenlines,epsfig,superscriptaddress]{revtex4}
\usepackage{graphicx}
\usepackage{color}
\usepackage[percent]{overpic}
\usepackage{enumerate}
\usepackage{enumerate}
\usepackage[export]{adjustbox}
\newcommand{\ket}[1]{|\,{#1}\,\rangle}
\newcommand{\braket}[2]{\mbox{$\langle\,{#1}\, | \,{#2}\,\rangle$}}

\newcommand{\expec}[1]{\langle #1 \rangle}

\newcommand{\sub}[2]{{#1}_{\mbox{\!\! \scriptsize #2}}}

\def\beq{\begin{equation}}
\def\eeq{\end{equation}}

\def\CR{\nonumber\\[0.15cm]}

\newcommand{\rref}[1]{Ref.~\cite{#1}}
\newcommand{\fref}[1]{Fig.~\ref{#1}}
\newcommand{\frefp}[2]{Fig.~\ref{#1}~(#2)}

\newcommand{\eref}[1]{Eq.~(\ref{#1})}
\newcommand{\esref}[2]{Eqs.~(\ref{#1}) and (\ref{#2})}

\newcommand{\sref}[1]{section~\ref{#1}}

\newcommand{\cref}[1]{chapter~\ref{#1}}

\newcommand{\Cref}[1]{Chapter~\ref{#1}}

\newcommand{\aref}[1]{appendix~\ref{#1}}
\newcommand{\bref}[1]{(\ref{#1})}

\usepackage{ulem}  %unter- (\uline{important}) und durchstreichen (\sout{wrong})
\begin{document}
%%%%%%%%%%%%%%%%%%%%%%%%%%%%%%%%%%%%%%%%%%%%%%%%%%%%%%%%%%%%%%%%%%%%%%%%

\title{Collisions of solitary waves in condensates beyond mean-field theory}
\author{Aparna Sreedharan}
\affiliation{Department of Physics, Indian Institute of Science Education and Research (IISER), Bhopal, Madhya Pradesh 462066, India}
\author{Sarthak Choudhury}
\affiliation{Department of Physics, Indian Institute of Science Education and Research (IISER), Bhopal, Madhya Pradesh 462066, India}
\author{Rick Mukherjee}
\affiliation{Department of Physics, Indian Institute of Science Education and Research (IISER), Bhopal, Madhya Pradesh 462066, India}
\affiliation{Department of Physics, Imperial College, SW7 2AZ, London, UK}
\author{Alexey Streltsov}
\affiliation{Theoretische Chemie, Physikalisch-Chemisches Institut, Universit{\"a}t Heidelberg, Im Neuenheimer Feld 229, D-69120 Heidelberg, Germany} \affiliation{SAP Deep Learning Center of Excellence and Machine Learning Research 
SAP SE, Dietmar-Hopp-Allee 16, 69190 Walldorf, Germany}
\author{Sebastian~W\"uster}
\affiliation{Department of Physics, Indian Institute of Science Education and Research (IISER), Bhopal, Madhya Pradesh 462066, India}
\email{sebastian@iiserb.ac.in}
\begin{abstract}
Bright solitary waves in a Bose-Einstein condensate contain thousands of identical atoms held together despite their only weakly attractive contact interactions. 
They nonetheless behave like a compound object, staying whole in collisions, with their collision properties strongly affected by inter-soliton quantum coherence.
We show that separate {solitary waves} decohere due to phase diffusion, dependent on their effective ambient temperature, after which their initial mean-field relative phases are no longer well defined or relevant for collisions. In this situation, collisions occur predominantly repulsively and can no longer be described within mean field theory. When considering the time-scales involved in recent {solitary wave} experiments where non-equilibrium phenomena play an important role, these features could explain the predominantly repulsive collision dynamics observed in most condensate soliton train experiments.
\end{abstract}
\maketitle
%
%%%%%%%%%%%%%%%%%%%%%%%%%%%%%%%%%%%%%%%%%%%%%%%%%%%%%%%%%%%%%%%%%%%%%%%%%%%%%%%%
\section{Introduction}
%%%%%%%%%%%%%%%%%%%%%%%%%%%%%%%%%%%%%%%%%%%%%%%%%%%%%%%%%%%%%%%%%%%%%%%%%%%%%%%%

Dilute alkali gas Bose-Einstein condensates (BECs) can usually be well understood using a simplified model for atomic collisions based on contact interactions and further employing a product mean-field Ansatz where all particles reside in the same single particle state to vastly simplify the  quantum many-body physics \cite{book:pethik,stringari:review}.

Here we explore why the mean-field approach breaks down in collisions of bright matter-wave solitary waves \cite{book:solitons,book:pethik,li_rev}, which are self-localized non-linear wave packets containing thousands of condensate atoms. Bright solitary matter-waves in Bose-Einstein condensates have now been created in a variety of experiments \cite{khay:brighsol,strecker:brighsol,li_rev,gap_exp,jila:solitons,Nguyen_modulinst,Nguyen_solcoll_controlled,Marchant_Quantrefl,Marchant_controlledform,Medley_evapsoliton,Lepoutre_sol_Ka,Boisse_disordersol_EPL,McDonald_solitoninterf,Everitt_modinst,Pollack_extreme_tunability_PRL}, for fundamental studies and applications in interferometry. We use ``Solitary wave" here to imply, that the three dimensional character of the wave function was still relevant in all these experiments. In the remainder of the article, we shall also use the shorthand ``soliton".
In many of these experiments, trains of $3-15$ solitons are created at once~\cite{strecker:brighsol,jila:solitons,Nguyen_modulinst,Pollack_extreme_tunability_PRL,Mesnarsic_cesiumsol_PhysRevA}, so that subsequently interactions or collisions between them become relevant. In mean-field theory, these should be akin to collisions of solitons in non-linear optics, which were well understood earlier \cite{gordon_forces}. Those results predict
 effectively attractive interactions for solitons with a mean-field relative phase of $\varphi=0$ and effectively repulsive interactions for out of phase solitons with $\varphi=\pi$.

Some early doubts were cast on these simple rules by a set of multi-soliton experiments (MSE), frequently commencing from explosively heated initial states. These indicated almost exclusively repulsive collisions \cite{strecker:brighsol,jila:solitons,Nguyen_modulinst}. However, a more controlled two-soliton experiment (TSE) shows collisions in apparent agreement with mean field theory \cite{Nguyen_solcoll_controlled,billam_NP}. While the MSE results could imply a robust creation of relative $\pi$ phases between all adjacent solitons \cite{stoof_solitons}, the creation of such a pattern cannot be accounted for by theory \cite{wuester:collsoll,brand_solitons,Salasnich_modinst}. Rather, studies beyond mean field theory reported dramatic modifications of soliton interactions by quantum effects \cite{wuester:collsoll,streltsov_frag}. 

Here we extend and consolidate the results of \cite{wuester:collsoll,streltsov_frag}, by identifying the two essential physical mechanisms that dynamically invalidate mean field theory. These are firstly phase diffusion \cite{lewenstein_phasediff} or loss of coherence between colliding solitons, and secondly atom transfer between solitons during a collision, akin to atom tunnelling in Bosonic-Josephon-Junction (BJJ) \cite{Albiez:oberthaler:BJJ}. The resultant picture provides more consistency with earlier experimental results than mean field theory.

We find that phase diffusion must lead to fragmentation of a train of solitons, which consequently exhibits more repulsive collision trajectories than it would otherwise. At zero temperature, the time scale for this fragmentation may be rather long, of the order of seconds. However, we show that fragmentation is significantly accelerated by thermal or uncondensed atoms, and thus can occur on ms time-scales for strongly heated condensates. 

This article is organized as follows: In  \sref{meanfield_solcoll}, we first review soliton collisions in mean-field theory and provide a brief overview of existing experiments on soliton trains and collisions. In  \sref{bmft}, we introduce the employed beyond-mean-field techniques. Using these techniques, we then first consider the fragmentation of non-interacting solitons in  \sref{phasediff}, and then move to the interplay of fragmentation and soliton collisions in  \sref{collisions}. This section separately considers collisions before fragmentation, \sref{before_frag}, after fragmentation,  \sref{after_frag}, the interplay with atom transfer during a collision \sref{sec_number_change}, see also appendix \ref{velocity_of_a}. {We discuss the relevance of broken many-body integrability during collisions in \sref{integrability} and the dependence on collision velocity in \sref{sec_velocity_dependence}.} We then move to a discussion of non-zero temperature and the ramifications of our results in the context of recent experiments in  \sref{disc_of_exp}. Finally, in  \sref{com}, we briefly compare the methods employed here, before concluding.

%%%%%%%%%%%%%%%%%%%%%%%%%%%%%%%%%%%%%%%%%%%%%%%%%%%%%%%%%%%%%%%%%%%%%%%%%%%%%%%%
\section{Mean-field soliton collisions}
\label{meanfield_solcoll}
%%%%%%%%%%%%%%%%%%%%%%%%%%%%%%%%%%%%%%%%%%%%%%%%%%%%%%%%%%%%%%%%%%%%%%%%%%%%%%%%

Let us first review soliton collisions in mean field theory. We consider a Bose gas with the second quantized Hamiltonian
\begin{align}
\label{Hamiltonian}
\nonumber \hat{H}&= \int d^3\mathbf{r} \bigg\{ \: \:\hat{\Psi}^\dagger
(\mathbf{r} )\left[-\frac{\hbar^2}{2m}\boldsymbol{\nabla}^2+ \frac{1}{2}m \omega_\perp^2 \mathbf{r}_\perp^2\right]\hat{\Psi}(\mathbf{r} )\\ &+ \frac{\sub{U}{3d}}{2} \hat{\Psi}^\dagger(\mathbf{r} )\hat{\Psi}^\dagger (\mathbf{r} )\hat{\Psi}(\mathbf{r} )\hat{\Psi}(\mathbf{r} ) \bigg\},
\end{align}
where the atomic field operator $\hat{\Psi}(\mathbf{r})$ destroys an atom of mass $m$ at location $\mathbf{r}$. The atoms experience 3D s-wave collisions with interaction strength $\sub{U}{3d}=4\pi\hbar^2 a_s/m$, where $a_s$ is the scattering length. The latter is controllable via a Feshbach resonance and assumed tuned to attractive interactions $a_s<0$ to enable bright solitons. Finally atoms are assumed free along the direction $x$, but
tightly trapped with trap-frequency $\omega_\perp$ in the transverse directions $\mathbf{r}_\perp=[y z]^T$.

In the simplest mean field treatment of \bref{Hamiltonian}, atomic quantum fluctuations are neglected and the field operator is replaced by the mean field condensate wave function $\phi(\mathbf{r})=\expec{\hat{\Psi}(\mathbf{r})}$.

We will be exclusively interested in the quasi one-dimensional (1D) scenario, where the Bose gas is more tightly confined along transverse coordinates $\mathbf{r}_\perp$ than in the longitudinal one $x$. We then implement the usual 1D reduction, where the 3D mean-field wave function is written as $\phi(\mathbf{r},t)=\phi(x,t)\eta(\mathbf{r}_\perp)$, with $\eta(\mathbf{r}_\perp)=\exp{[-\mathbf{r}_\perp^2/(2\sigma_{\perp}^2)]}/(\sigma_{\perp}\sqrt{\pi})$, thus assuming transversally the BEC remains in the trap ground-state with width $\sigma_{\perp}=\sqrt{\hbar/(m\omega_\perp)}$, using the transverse oscillator frequency $\omega_\perp$. Then we can derive the quasi-1D Gross-Pitaevskii equation (GPE) for the evolution of the longitudinal mean-field $\phi(x,t)$ using standard methods as
\begin{align}
\label{GPE}
i\hbar \frac{\partial}{\partial t}\phi(x,t) =  \left[ -\frac{\hbar^2}{2m} \frac{\partial^2}{\partial x^2}  + U_0{|\phi(x,t)|}^2\right] \phi(x,t), 
\end{align}
where $U_{0}=\sub{U}{3d}/(2\pi\sigma_{\perp}^2)$ is the effective interaction strength. Importantly, this does not imply that microscopic collisions are constrained to 1D.

\begin{figure}[htb]
\includegraphics[width=0.99\columnwidth]{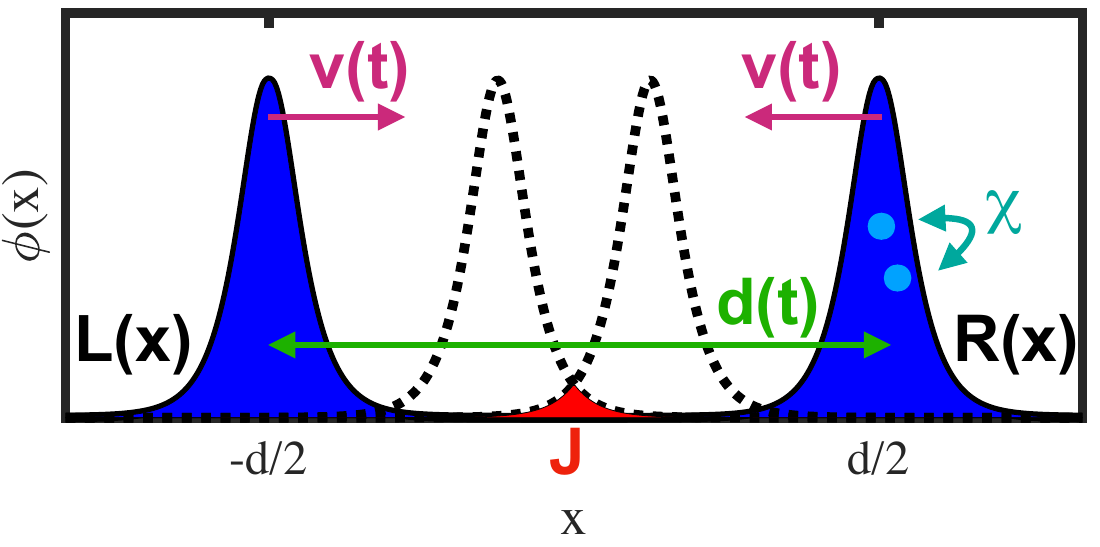}
\caption{\label{sketch} Sketch of colliding soliton pair with mean field wave function \bref{soliton_pair_meanfield}. During the collision, the separation $d(t)$ and velocity $v(t)$ evolve in time. We also sketch two causes for the break down of mean field theory: Phase diffusion due to atom-atom interactions within a soliton $\chi$, and atom transfer between solitons $J$, when they are in close proximity.}
\end{figure}

For attractive interactions $U_0<0$, the GPE \bref{GPE} has stationary soliton solutions $\phi(x,t) = \sub{\phi}{sol}(x)\exp{[-i\mu t/\hbar]}$
with a spatial profile as sketched in \fref{sketch}
\begin{align}
\label{soliton}
\sub{\phi}{sol}(x) = {\cal N}\:\: \mbox{sech}(x/\xi),
\end{align}
where $\mu=-m \sub{N}{sol}^2 U_0^2/8/\hbar^2$ is the chemical potential if the soliton contains $\sub{N}{sol}$ atoms, ensured by the normalisation factor ${\cal N}=\sqrt{2 | \mu |/U_0}$. The width of the soliton is set by the healing length $\xi=\sqrt{\hbar^2/(2m | \mu |)}$. While the solution \bref{soliton} is strictly valid for a 1D system only, it aptly describes bright condensate solitons in realistic 3D experiments as long as the transverse trapping $\omega_{\perp}$ is sufficiently tight \cite{parker_bsw} and $\sub{N}{sol}$ remains safely away from the critical atom number $\sub{N}{crit}$ for 3D collapse \cite{jila:solitons}, which we do not discuss here.

To study soliton collisions, we now move to a mean-field wave function containing a pair of solitons
\begin{align}
\phi(x,t) &=L(x,t)e^{ik(t)x} +e^{i\varphi(t)} R(x,t)e^{-ik(t)x},
\label{soliton_pair_meanfield}
\end{align}
with left and right soliton shapes $L(x,t)=\sub{\phi}{sol}(x+d(t)/2)$, $R(x,t)=\sub{\phi}{sol}(x-d(t)/2)$. The two solitons are thus separated by a distance $d$. We also allow a wave number 
$k$ arising from symmetric soliton motion. The Ansatz \bref{soliton_pair_meanfield} is sketched in \fref{sketch}. For a simplified description, 
we can use a time-dependent variational principle from the Lagrangian based on \bref{Hamiltonian} \cite{gordon_forces,stoof_solitons}, to derive the effective kinetic equations of motion
\begin{align}
\frac{\partial^2}{\partial t^2} \varphi(t) &=8 \exp{[-d(t)]}\sin{[\varphi(t)]},
\label{soliton_phase_eom}
\\
\frac{\partial^2}{\partial t^2} d(t) &= -8 \exp{[-d(t)]}\cos{[\varphi(t)]},
\label{solitoneom}
\end{align}
for the time evolving soliton separation $d(t)$, velocity $v(t) = \hbar k(t)/m$ and relative phase $\varphi(t)$, still based on mean field theory. We write \bref{soliton_phase_eom}-\bref{solitoneom} for dimensionless variables with $\xi=1$, $\hbar=1$, $m=1$. Clearly for $\varphi(t=0)=0$ ($\pi$), the relative phase does not evolve.
We further see that a relative phase $\varphi=0$ yields attractive and $\varphi=\pi$ repulsive behaviour \cite{gordon_forces,stoof_solitons}. 

\begin{figure}[htb]
\includegraphics[width=0.99\columnwidth]{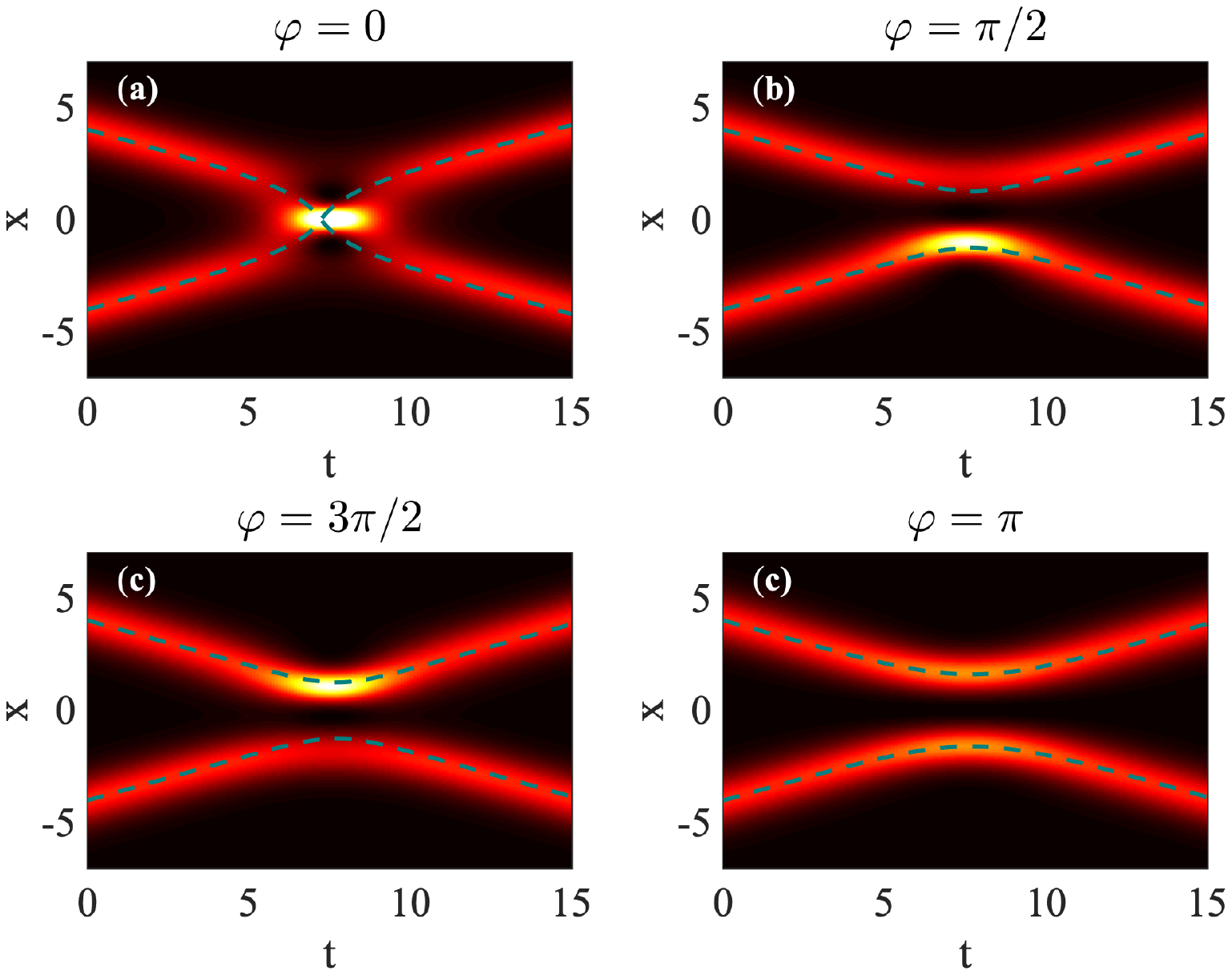}
\caption{\label{gpecoll} Review of soliton collisions in mean-field theory for initial relative phase $\varphi$ of (a) $\varphi=0$, (b) $\varphi=\pi/2$, (c) $\varphi=3\pi/2$ and (d) $\varphi=\pi$. The color-profile is the atomic density $|\phi(x,t)|^2$ following from \bref{GPE} for the initial state \bref{soliton_pair_meanfield} (black-zero, bright-high). Overlaid dashed teal lines are trajectories based on \bref{solitoneom}.}
\end{figure}

We illustrate in \fref{gpecoll} that the effective kinetic equations \bref{soliton_phase_eom}-\bref{solitoneom} indeed correctly reproduce soliton dynamics predicted by the GPE \bref{GPE}.
%  

%%%%%%%%%%%%%%%%%%%%%%%%%%%%%%%%%%%%%%%%%%%%%%%%%%%%%%%%%%%%%%%%%%%%%%%%%%%%%
\subsection{Experiments with soliton trains and collisions}
\label{sec_experiments}
%%%%%%%%%%%%%%%%%%%%%%%%%%%%%%%%%%%%%%%%%%%%%%%%%%%%%%%%%%%%%%%%%%%%%%%%%%%%%%%%

While \eref{solitoneom} was largely verified in non-linear optics relatively soon after its prediction \cite{mitschke_forces}, it is still not fully clear to what extent or under which conditions it describes matter-wave solitons.

Soon after the first creation of single matter-wave solitons \cite{khay:brighsol}, experiments began to investigate trains or collections of multiple solitons~\cite{jila:solitons,strecker:brighsol}, that appear when the interactions within a large 1D BEC cloud are suddenly changed from repulsive to attractive using a Feshbach resonance. This led to condensate collapse with strong loss of atoms and heating, with remnant atoms forming a train of solitons.
 Both experiments saw indirect evidence for dominantly repulsive interactions between neighbouring solitons in the train: (i) the total remnant atom number after the collapse was still higher than the critical atom number $\sub{N}{crit}$ \cite{ruprecht:attractive} for further collapse, (ii) soliton trajectories, within limited experimental resolution, were typically repulsive, (iii) almost all solitons survive collisions, which is not the case when interactions are attractive due to further 3D collapse \cite{parker_previouspreprint,parker_bsw1}. A relative soliton phase of $\pi$ between neighbors would explain this behavior~\cite{li_rev,stoof_solitons, li_theory} as discussed in \sref{meanfield_solcoll}, but such a phase pattern should not actually arise, according to theory \cite{wuester:collsoll,brand_solitons,Salasnich_modinst}. A striking counter example are the repulsively interacting soliton trains with two and four members in~\cite{jila:solitons}, for which symmetry requires a phase  $\pi=0$ between the central two solitons.
 
To address these questions, among others, further experiments were recently performed in the Rice group, tracking soliton collisions using in situ observation. In one case, \cite{Nguyen_solcoll_controlled}, a condensate was split into two pieces, which were subsequently transformed into solitons in a fairly controlled process, which however took much longer than condensate collapse. We refer to \rref{Nguyen_solcoll_controlled} as the two-soliton experiment (TSE) later. More recently, another soliton-train was studied resulting from modulational instability \cite{Nguyen_modulinst}, in a multi-soliton-experiment (MSE). In contrast to earlier MSE~\cite{jila:solitons,strecker:brighsol}, a  violent initial 3D collapse of the the entire cloud into essentially one single high density spike was avoided. The TSE demonstrated that multiple collisions of one soliton pair can be described by the GPE or \eref{solitoneom}, \emph{provided that} their relative phase is in fact inferred indirectly from the first of those collisions. 
The MSE in turn, again found necessarily repulsive interactions between all neighboring solitons of trains with up to $10$ (even) members, since their number remained constant despite the fact that attractive interactions should have resulted in 3D collapse.

It will be important for this article, that none of the experiments discussed were in a genuinely 1D regime where one can view also microscopic collisions between atoms as constrained to one dimension. Rather they all fall into the quasi-1D regime, where the BEC system can be reliably mathematically modelled taking into account only one dimension in the trap, but microscopic atomic collisions would still significantly involve three dimensions.

In the following we combine earlier indications of beyond mean-field effects in soliton collisions \cite{wuester:collsoll,streltsov_frag} to develop a more comprehensive picture that can reconcile most experimental results above and additionally are suggestive of further quantum dynamical effects, such as entanglement generation, as subject for future experiments. 

%%%%%%%%%%%%%%%%%%%%%%%%%%%%%%%%%%%%%%%%%%%%%%%%%%%%%%%%%%%%%%%%%%%%%%%%%%%
\section{Beyond mean-field theories}
\label{bmft}
%%%%%%%%%%%%%%%%%%%%%%%%%%%%%%%%%%%%%%%%%%%%%%%%%%%%%%%%%%%%%%%%%%%%%%%%%%%%%%%%

As discussed above, there are experimental and theoretical indications, that collisions of bright matter wave solitons may be a case where mean-field theory suffers not only a quantitative but a qualitative break down. In this section we now briefly summarize three different beyond mean-field models that can explore this aspect. 

%%%%%%%%%%%%%%%%%%%%%%%%%%%%%%%%%%%%%%%%%%
\subsection{Two-mode model} 
\label{tmm}
%%%%%%%%%%%%%%%%%%%%%%%%%%%%%%%%%%%%%%%%%%

One way to go beyond the mean field expression \bref{soliton_pair_meanfield}, is with a simple two-mode-model (TMM) for the field operator
\begin{align}
\label{twomode}
\hat{\Psi}(x) =\overline{L}(x) \hat{a} +\overline{R}(x)   \hat{b},
\end{align}
where the left and right ``soliton mode functions`` $\overline{L}(x) =L(x)/\sqrt{\sub{N}{sol}}$ and $\overline{R}(x) =R(x)/\sqrt{\sub{N}{sol}}$ are now normalized to one instead of $\sub{N}{sol}$ but retain the shape of soliton. The operator $\hat{a}$ destroys a boson in the left soliton and  $\hat{b}$ does the same for the right soliton, they act on Fock states $\ket{n,m}$, where $n$ ($m$) is the number of atoms in the left (right) soliton. Thus atomic spatial degrees of freedom are constrained to reside in either the left or right soliton mode. In principle the TMM can straightforwardly be extended to describe the problem in three spatial dimensions, by augmenting to modes $\overline{L}(\mathbf{x})$, $\overline{R}(\mathbf{x})$ to 3D functions, assuming the same simplified transverse shape as discussed in \sref{meanfield_solcoll}.

We now allow number fluctuations, and through these, varying phase relations. In a Fock state $\ket{n,m}$ the relative phase between solitons is undefined, while in a two mode coherent state 
\begin{align}
\ket{\alpha,\beta}=\ket{\alpha}\otimes\ket{\beta}
\label{twomodecohst}
\end{align}
with $\ket{\alpha}=e^{-\frac{|\alpha|^2}{2}} \sum_{n=0}^\infty \frac{\alpha^n}{\sqrt{n!}}\ket{n}$ the relative phase is $\varphi = \mbox{arg}[\alpha]- \mbox{arg}[\beta]$. This two mode coherent state has an uncertain total atom number.

Even for fixed total atom number $\sub{N}{tot}$ we can assign a well defined relative phase between left and right soliton, using a relative coherent state in the even or odd soliton pair $\ket{\sub{N}{tot},\pm}\equiv [(\hat{a} \pm \hat{b})/\sqrt{2}]^{\sub{N}{tot}}/\sqrt{\sub{N}{tot}!}\ket{0}$.

Inserting \bref{twomode} into \bref{Hamiltonian}, assuming \emph{real} mode functions with a narrow Gaussian shape along transverse directions, we obtain the TMM Hamiltonian
\begin{align}
\hat{H} &= \omega( \hat{a}^\dagger \hat{a} + \hat{b}^\dagger \hat{b} )+  \frac{\chi}{2} ( \hat{a}^\dagger \hat{a}^\dagger \hat{a} \hat{a} + \hat{b}^\dagger \hat{b}^\dagger \hat{b} \hat{b}) 
\CR
&+J (\hat{b}^\dagger \hat{a} + \hat{a}^\dagger \hat{b}) +  \bar{U}(4 \hat{a}^\dagger \hat{a} \hat{b}^\dagger  \hat{b}+ \hat{a}^\dagger \hat{a}^\dagger \hat{b} \hat{b} +\hat{b}^\dagger \hat{b}^\dagger \hat{a} \hat{a} ) \CR
&+2\bar{J}(\hat{a}^\dagger \hat{a} + \hat{b}^\dagger \hat{b}  -1 )(\hat{b}^\dagger \hat{a} + \hat{a}^\dagger \hat{b}),
\label{twomodeHamil}
\end{align}
with coefficients
\begin{subequations}
\begin{align}
\omega&=\int dx \:  \bar{L}(x)\left[-\frac{\hbar^2}{2m} \frac{\partial^2}{\partial x^2}\right]\bar{L}(x),\\
\chi&=U_0\int dx \: \bar{L}(x)^4=-\frac{m U_0^2 \sub{N}{sol}}{6 \hbar^3},
\label{chidef}
 \\
J(d)&=\int dx \:  \bar{L}(x)\left[-\frac{\hbar^2}{2m} \frac{\partial^2}{\partial x^2}\right]\bar{R}(x),\\
 \bar{U}(d)&=\frac{U_0}{2}\int dx  \: \bar{L}(x)^2 \bar{R}(x)^2, \\
\bar{J}(d)&=\frac{U_0}{2}\int dx  \:\bar{L}(x)^3 \bar{R}(x).
\end{align}
\label{twomodecoeffsl}
\end{subequations}
We indicated with an argument $d$ whether a coefficient depends on the distance between the left and right solitons modes.
The TMM will be useful in \sref{phasediff} to elucidate the basic physics underlying the predictions of the more involved quantum many-body theories discussed further below.

For large $d$, when $J,\bar{U},\bar{J}\rightarrow0$, the TMM can be analytically solved, as shown in \sref{phasediff}. In the more general case, we will numerically solve the time-dependent Schr{\"o}dinger equation for $\ket{\Psi(t)}$, \emph{coupled} to \eref{solitoneom} via $d(t)$. The coefficients $J$, $\bar{U}$, $\bar{J}$ in \eref{twomodeHamil} then vary in time, due to their dependence on the soliton separation $d(t)$. We used $\ket{\alpha,\beta}$ as initial-state when comparing with TWA (see \sref{twa}) and $\ket{\sub{N}{tot},\pm}$ for comparisons with MCTDHB (see \sref{mctdhb}) .

%%%%%%%%%%%%%%%%%%%%%%%%%%%%%%%%
\subsection{Multi-configurational time-dependent Hartree for Bosons (MCTDHB)} 
\label{mctdhb}
%%%%%%%%%%%%%%%%%%%%%%%%%%%%%%%%

The TMM is made more sophisticated in MCTDHB \cite{alon:pra:mctdhb} with two orbitals. In essence the latter allows a combination of the mean-field and the two-mode approach. It  allows the bosons to condense into two orbitals, as the quantum field operator is again expanded as:
\begin{align}
\label{mcfieldop}
\hat{\Psi}(x,t) =\phi_+(x,t) \hat{c}(t) + \phi_-(x,t)  \hat{d}(t).
\end{align}
This includes the Ansatz \bref{twomode} but importantly now contains two orbitals $\phi_\pm(x,t)$ that can self-consistently evolve in time. Their evolution and that of the Fock states onto which $\hat{c}$, $\hat{d}$ act is determined from a time-dependent many-body variational principle \cite{alon:pra:mctdhb}. In contrast, in the TMM, the time-dependence of $L$, $R$ is fixed a priori. 

Initially, the orbitals are taken as the symmetric or anti-symmetric linear combination of the soliton modes $\phi_\pm(x,t)=[\bar{L}(x) \pm \bar{R}(x)]/\sqrt{2}$. Depending on the initial relative phase $\varphi$ between the solitons, either is initially fully occupied. Since MCTDHB operates with a fixed total atom number, the corresponding initial state in the two-mode model is $\ket{\sub{N}{tot},\pm}$.

We refer to the original article \cite{alon:pra:mctdhb} for the equations of motion, and the extensive literature for details. The method has proven particularly useful to study scenarios involving dynamical condensate fragmentation \cite{Streltsov_fragmentation_PRL2008,alon:attractive:anharmonic,Sakmann_universalfrag_PhysRevA,Katsimiga_impurity_PRA,Katsimiga_bentdark_NJP2017}, scenarios generating entanglement \cite{Streltsov_barrier_PRA2009,Grond_Shapiro_NJP2011,Kroenke_impurity_NJP2015,Katsimiga_darkbright_NJP2017} and few-body dynamics \cite{cosme_intblock_pra}. 
Here we use the open-MCTDHB package \cite{open_mctdhb}.

Important approximations contained in the Ansatz \ref{mcfieldop} are the reduction from three spatial dimensions to one, with the same argumentation as in \sref{meanfield_solcoll}, and the reduction of the many-mode quantum field problem to the substantially simplified two mode constraint.

%%%%%%%%%%%%%%%%%%%%%%%%%%%%%%%%
\subsection{Truncated Wigner Approximation (TWA)} 
\label{twa}
%%%%%%%%%%%%%%%%%%%%%%%%%%%%%%%%
%
We finally drop the two-mode constraint, moving to an (approximate) multi-mode quantum field theory. An effective approximation technique for those is the truncated Wigner framework \cite{steel:wigner,Sinatra2001,castin:validity,blair:review}, where the quantum many body state is represented by an ensemble of stochastic trajectories. In TWA we solve the same equation of motion as for mean field theory \bref{GPE}, albeit with random noise added to the initial state 
\begin{align}
\label{TWAnoise}
\phi(x,0)=\phi_0(x) + \sum_{\ell} \frac{\eta_{\ell}u_\ell(x)}{\sqrt{2 \tanh{( \frac{\epsilon_{\ell}}{2k_{B}T})}}},
\end{align}
with $\phi_0(x)$ the mean field soliton pair \bref{soliton_pair_meanfield}. The index $\ell$ numbers a plane wave basis $u_\ell(x)=e^{i k_\ell x}/{\sqrt{\cal V}}$ with normalisation volume ${\cal V}$, then $\epsilon_{\ell}= \hbar^2 k_\ell^2/(2m)$. The $\eta_{\ell}$ are complex Gaussian noises with unit variance and correlations $\overline{\eta_{\ell}\eta_{j}}=0$, $\overline{\eta_{\ell}^*\eta_{j}}=\delta_{\ell j}$ and $T$ is the system temperature. Overlines indicate stochastic averages. The TWA described here is known to give good results for decoherence phenomena \cite{hush_numberphasewigner,corney_nongauss_posp_wig} as long as the noise amplitude added is dominated by the meanfield \cite{polkovnikov:timescale,norrie:prl,norrie:long,norrie:thesis}, but it would usually fail to capture e.g.~quantum revivals \cite{hush_numberphasewigner,corney_nongauss_posp_wig} such as exhibited by the model \bref{twomodeHamil} at later times.

Quantum correlations are extracted according to 
\begin{align}
\label{averagesTWA}
\expec{\hat{\Psi}^\dagger(x') \hat{\Psi}(x)} &= \overline{\phi^*(x')\phi(x)}-\frac{1}{2}\delta_c(x,x'),
\end{align}
where $\delta_c(x,x')=\sum_\ell u_\ell(x)u_\ell^*(x')$ is a restricted basis commutator \cite{norrie:thesis}. Our TWA calculations and TMM solutions employ the XMDS package \cite{xmds:docu,xmds:paper}. We will later use TWA for comparison with MCTDHB, simulation of experiments and for the incorporation of finite temperature.

The truncated Wigner method derives it name from the truncation of the evolution equation for the distribution function of stochastic trajectories, in order to bring that into the form of a Fokker-Planck equation \cite{steel:wigner}. This approximation is motivated by practicalities and its physical implications often far from obvious. Later work has however shown, that the approximation will be good for short times (where it is again not a-priori obvious how short) \cite{polkovnikov:timescale} and as long as most modes of the quantum field are highly occupied \cite{Sinatra2001,castin:validity}. In practice this means that the noise amplitude added in \bref{TWAnoise} ought to be small compared to the mean-field amplitude $|\phi_0(x)|$ \cite{norrie:prl,norrie:long,norrie:thesis}.

%%%%%%%%%%%%%%%%%%%%%%%%%%%%%%%%
\subsection{Coherence and Fragmentation} 
%%%%%%%%%%%%%%%%%%%%%%%%%%%%%%%%
%
Within all three many-body models, we are mainly interested in the resultant coherence and fragmentation dynamics. To identify the condensate 
in a quantum-field setting, we use the Penrose-Onsager criterion \cite{penrose_onsager_crit,book:pethik,Blakie2005}, that the largest eigenvalue of the one-body density matrix (OBDM) is the condensate occupation, with OBDM
\begin{align}
\label{obdm}
\varrho(x,x')=\expec{\hat{\Psi}^\dagger (x')\hat{\Psi} (x)}.
\end{align}
The eigenvalues $\lambda_j $ are then obtained from $\int dx'\varrho(x,x')\chi_j(x') = N \lambda_j \chi_j(x')$ where $\chi_j(x)$ is the corresponding single particle orbital and $N=2\sub{N}{sol}$. If two $ \lambda_j $ are of order unity, the system is called \emph{fragmented} \cite{book:pethik}.
In the TWA the OBDM is given by \bref{averagesTWA}, and in MCTDHB by $\varrho(x,x')=\sum_{kq} \expec{\hat{O}_k^\dagger\hat{O}_q} \phi^* _k(x',t)\phi_q(x,t)$ \cite{streltsov:triplewellMCTDHB:PhysRevA.83.043604}, using $k,q\in\{+,-\}$ and $\hat{O}_+=\hat{c}$, $\hat{O}_-=\hat{d}$, see \bref{mcfieldop}.

For the TMM, we can ignore the frozen spatial structure and focus on the mode space OBDM 
\begin{align}
\label{modes_obdm}
\varrho= \begin{bmatrix}
\expec{\hat{a}^\dagger\hat{a}}  & \expec{\hat{b}^\dagger\hat{a}} \\
\expec{\hat{a}^\dagger\hat{b}}   & \expec{\hat{b}^\dagger\hat{b}} 
\end{bmatrix}.
\end{align}
We denote the two eigenvalues of $\varrho$ with $\lambda_+$ (the larger one) and  $\lambda_-$ (the smaller ones), in the following. 

%%%%%%%%%%%%%%%%%%%%%%%%%%%%%%%%%%%%%%%%%%%%%%%%%%%%%%%%%%%%%%%%%%%%%%%%%%%%%%%%
\section{Soliton pair fragmentation due to phase diffusion}
\label{phasediff}
%%%%%%%%%%%%%%%%%%%%%%%%%%%%%%%%%%%%%%%%%%%%%%%%%%%%%%%%%%%%%%%%%%%%%%%%%%%%%%%%

\begin{figure}[htb]
\includegraphics[width=0.79\columnwidth]{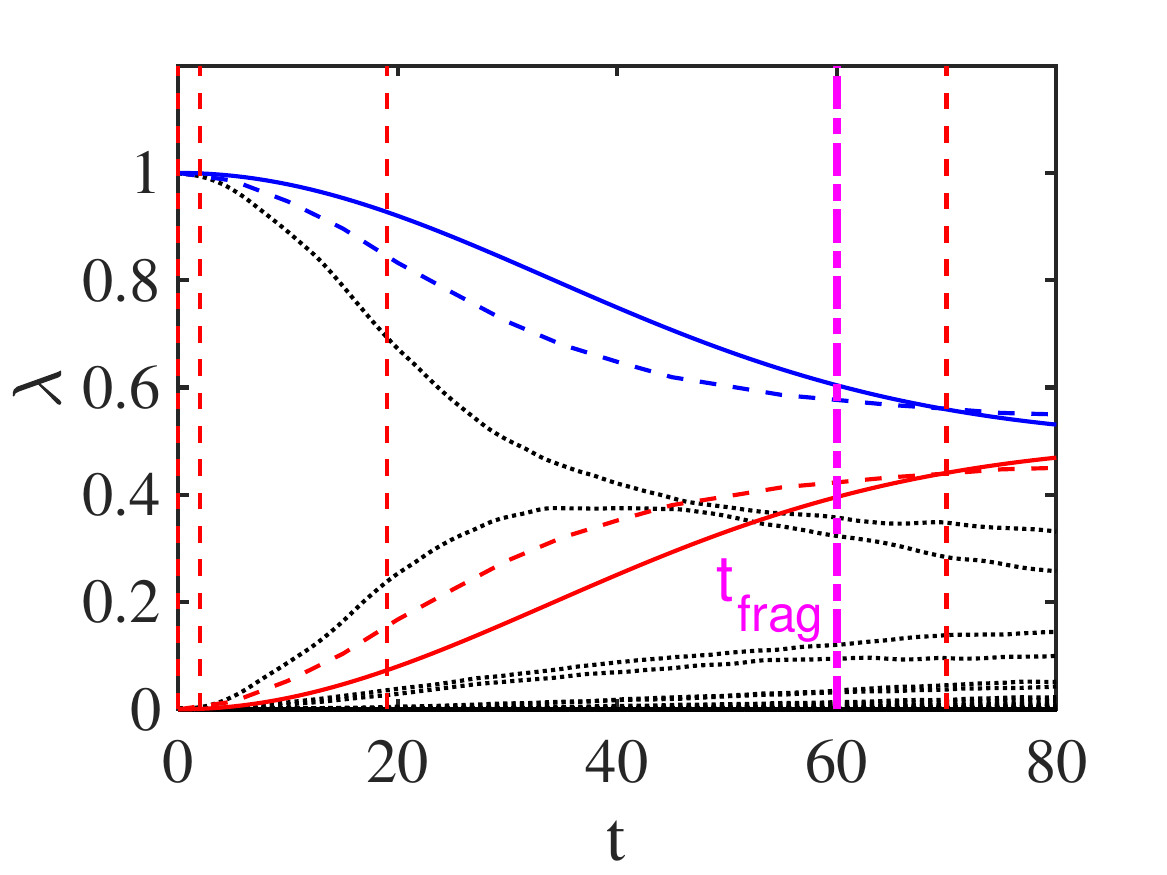}
\caption{\label{fragmentation_allmethods}
Fragmentation of far separated BEC solitons. We show the relative occupation $\lambda$ of all system orbitals at zero temperature in TWA (black dotted), MCTDHB (blue  and red dashed) and two-mode model (blue and red solid).  For MCTDHB and two-mode model there are only two orbitals by construction, for TWA two dominate. Initially we have a pure BEC of two solitons since $\lambda=1$ for one orbital. 
It then fragments around $\sub{t}{frag}$ as defined in the text, indicated by the vertical magenta dot-dashed line. The non-linear parameter, see \eref{chidef}, is $\chi=-6.6\times 10^{-4}$. Vertical red-dashed lines are the times for which we plot snapshots of the Q-function in \fref{phase_diffusion}. }
\end{figure}

We now initially consider the beyond mean-field evolution of two solitons far separated from each other so that they can be considered non-interacting.
They are initialized as part of one coherent, non-fragmented BEC. We show in \fref{fragmentation_allmethods} the eigenvalues of the OBDM predicted by all three methods discussed above.

It is clear that by the indicated time $\sub{t}{frag}$ eigenvalues $\lambda_+$ and  $\lambda_-$ have become comparable and the system is thus fragmented. We formally call the system fragmented after $\sub{t}{frag}$, when $|\lambda_+(\sub{t}{frag})-\lambda_-(\sub{t}{frag})|\equiv\Delta \lambda = 0.2$. The choice of $\Delta \lambda$ is somewhat arbitrary. We cannot chose $\Delta \lambda=0$, since we later show cases where the $\lambda_\pm$ are never quite equal, yet get very close and should still indicate fragmentation.

All three methods agree on the fragmentation time-scale defined above. Quantitative differences are expected, due to the varying numbers of modes and constraints on these among the methods.
The origin of fragmentation is best understood in the TMM. The coefficients $J$, $\bar{U}$, $\bar{J}$ in \bref{twomodeHamil} depend on the overlap of $\bar{L}(x)$ and $\bar{R}(x)$ and thus on $d$.
For large soliton separations $d$, all these vanish, and only the first line in \bref{twomodeHamil} remains. The dynamics can then be determined analytically 
\begin{align}
\ket{\Psi(t)}&=\sum_{nm}c_{nm}(t)\ket{nm},  \CR
c_{nm}(t)&=c_{nm}(0)e^{-i\left[E_0(n+m) + \frac{\chi}{2} (n(n-1)+m(m-1)) \right] t/\hbar},
\label{twomodesolution}
\end{align}
where the coefficients $c_{nm}(0)$ are set by the two-mode coherent initial state \bref{twomodecohst} with amplitude $\alpha,\beta=\sqrt{\sub{N}{sol}}$.
From \bref{twomodesolution} we obtain the eigenvalues of \bref{modes_obdm} as 
\begin{align}
\lambda_{\pm}&=\frac{1  \pm e^{2 \sub{N}{sol} [\cos(\chi t/\hbar)-1]}}{2}   \approx \frac{1\pm e^{-(t/\sub{t}{frag})^2}}{2},
\label{eigenvalues}
\end{align}           
where the expression after $\approx$ is valid for short times. The fragmentation timescale $\sub{t}{frag}=\hbar/( \sqrt{\sub{N}{sol}}|\chi|)$ is corroborated by the more involved quantum many body methods TWA and MCTDHB in \fref{fragmentation_allmethods}. For the TWA results in \fref{fragmentation_allmethods}, we can see the emergence of several additional significantly occupied orbitals beyond the first two. We will comment on these in \sref{collisions}.

Note that the Hamiltonian \bref{twomodeHamil} for large $d$ reduces to $\hat{H}=\frac{\chi}{2} ( \hat{a}^\dagger \hat{a}^\dagger \hat{a} \hat{a} + \hat{b}^\dagger \hat{b}^\dagger \hat{b} \hat{b}) $, after we adjust the zero of energy such that the term $\sim \omega$ can be ignored. This just corresponds to two independent non-linear Kerr oscillators and the dynamics just discussed thus is well known and referred to as Kerr-squeezing \cite{book:walls:milburn,matthias:simon:kerr,wuester:kerr} or phase diffusion \cite{lewenstein_phasediff}. Phase-diffusion refers to an initially fixed condensate mean phase becoming ill defined due to diffusion over all angles.

\begin{figure}[htb]
\includegraphics[width=0.99\columnwidth]{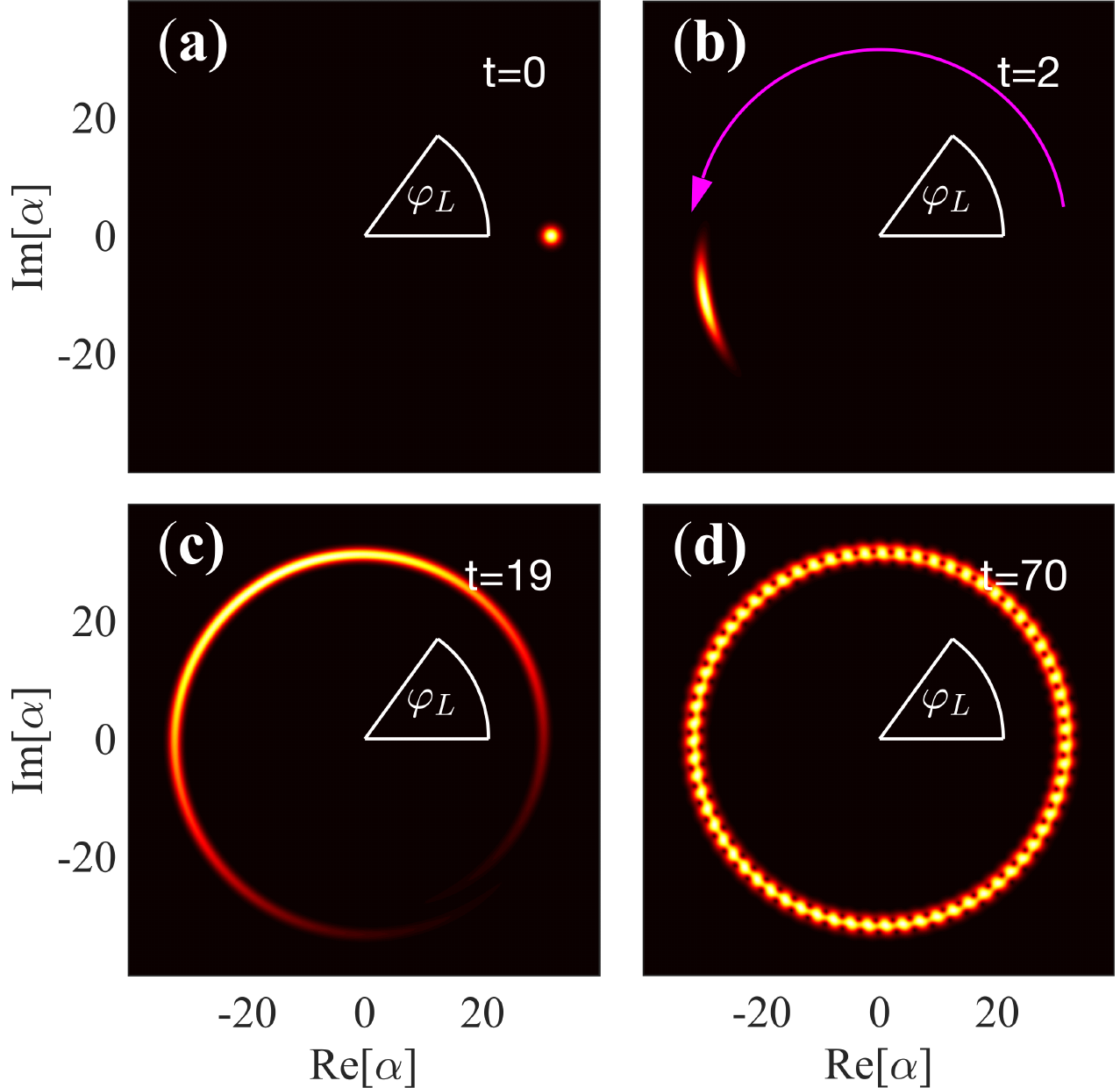}
\caption{\label{phase_diffusion} Phase diffusion  for the same case as in \fref{fragmentation_allmethods}, at times indicated there as red-dashed vertical lines. We plot the 
Husimi Q-function $Q(\alpha)$ of a single soliton's internal state \cite{book:walls:milburn}, see text. (black-zero, bright-high) (a) Initially, $t=0$, this corresponds to a coherent state. For later times as indicated (b,c,d), 
the Q-function shows shearing (Kerr-squeezing) and eventually indicates a completely undefined soliton phase $\varphi_L$.
 }
\end{figure}

We visualize phase diffusion for the reduced state of just one (the left) soliton in \fref{phase_diffusion}, using the Husimi Q-function $Q(\alpha)=|\braket{\alpha}{\Psi}|/\pi$ that quantifies the overlap of an arbitrary state $\ket{\Psi(t)}$ with a coherent state $\ket{\alpha}$. In the space $\alpha\in \mathbb{C}$, farther from the origin corresponds to larger atom number $n$ in the left soliton, and the argument of $\alpha$ indicates the soliton phase $\varphi_L$. 
We show $Q(\alpha)$ at several characteristic snapshots, indicated in \fref{fragmentation_allmethods} by vertical red dashed lines. 
Initially, the state of atom number within one of the solitons itself is a coherent state, with a 2D Gaussian as Q-function. It then shears, since the angular phase evolution due to non-linear interactions scales as 
$\varphi_L\sim \chi n(n-1)t$ with the atom number, and it thus faster for $\alpha$ farther away from the origin. During this initial period, see e.g.~$t=2$, the dynamics is also called Kerr squeezing.
At later times, the phase of a single soliton, and hence even more so the relative phase between two solitons becomes progressively undefined. At that stage, there also is complete fragmentation. 

Phase diffusion in the context of BEC solitons was explored before \cite{lai_quantsol_I,lai_quantsol_II}. It has been linked to fragmentation in the context of soliton interferometry \cite{Martin_solinterf_NJP_2012}. Here we clearly identify it as the root physical cause of soliton train fragmentation, first reported in \cite{streltsov_frag}. Most importantly, this enables us to make analytic predictions for the fragmentation time-scale $\sub{t}{frag}$ in \bref{eigenvalues} and will in the future allow assessments how fragmentation would depend on the number statistics of the initial state.

%%%%%%%%%%%%%%%%%%%%%%%%%%%%%%%%%%%%%%%%%%%%%%%%%%%%%%%%%%%%%%%%%%%%%%%%%%%%%%%%
%%%%%%%%%%%%%%%%%%%%%%%%%%%%%%%%%%%%%%%%%%%%%%%%%%%%%%%%%%%%%%%%%%%%%%%%%%%%%%%%
\section{Soliton collisions}
\label{collisions}
%%%%%%%%%%%%%%%%%%%%%%%%%%%%%%%%%%%%%%%%%%%%%%%%%%%%%%%%%%%%%%%%%%%%%%%%%%%%%%%%
%%%%%%%%%%%%%%%%%%%%%%%%%%%%%%%%%%%%%%%%%%%%%%%%%%%%%%%%%%%%%%%%%%%%%%%%%%%%%%%%

We now consider the effect of the fragmentation discussed above on the collisions of condensate solitons. We want to distinguish two cases, collisions occuring before fragmentation and after fragmentation. To this end initially un-fragmented solitons separated by a distance $\sub{d}{ini}$ are given an initial velocity $\sub{v}{ini}$ towards each other such that their expected collision time is approximately $\sub{t}{coll}=|\sub{d}{ini}/(2\sub{v}{ini})|$. We show in \fref{collisions_prior} and \fref{collisions_post} the atom density in a colliding soliton pair from MCTDHB (\sref{mctdhb}) as color shade, compared with the collision trajectory based on the kinetic equation \bref{solitoneom} as overlayed dashed teal line. Since \bref{solitoneom} is based on the GPE \bref{GPE}, we are thus directly comparing mean-field with beyond-mean-field collisions.

In both figures, the collision velocity is adjusted to $\sub{v}{ini}=0.2$. We then employ $\sub{d}{ini}=8$ in \fref{collisions_prior}, yielding an expected collision time $\sub{t}{coll}=20$, which is \emph{before} the expected fragmentation time of $\sub{t}{frag}=60$ for far separated solitons, based on \eref{eigenvalues}. 
In contrast, for \fref{collisions_post}, $\sub{d}{ini}$ is changed to $\sub{d}{ini}=32$ , hence $\sub{t}{coll}=80$  becomes larger than $\sub{t}{frag}$.

%%%%%%%%%%%%%%%%%%%%%%%%%%%%%%%%%%%%%%%%%%%%%%%%%%%%%%%%%%%%%%%%%%%%%%%%%%%%%%%%
\subsection{Before fragmentation}
\label{before_frag}
%%%%%%%%%%%%%%%%%%%%%%%%%%%%%%%%%%%%%%%%%%%%%%%%%%%%%%%%%%%%%%%%%%%%%%%%%%%%%%%%

\begin{figure}[htb]
\includegraphics[width=0.99\columnwidth]{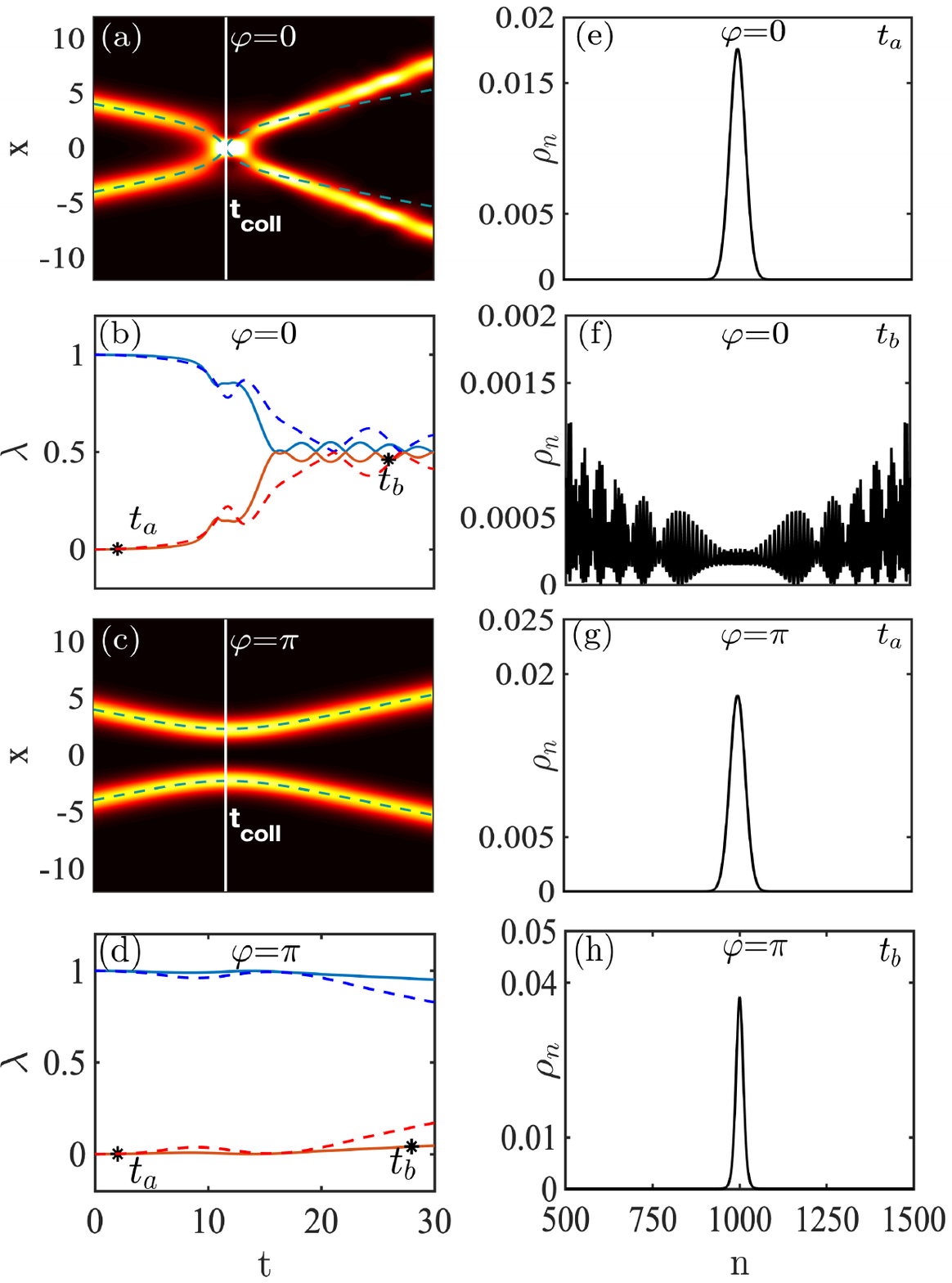}
\caption{\label{collisions_prior} Collision and coherence dynamics in controlled soliton collisions \emph{before} fragmentation, $\sub{t}{coll}<\sub{t}{frag}$. The initial relative phases between solitons, $\varphi$, are indicated. (a,c) Total atomic density (black-zero, bright-high) from MCTDHB and expected mean-field trajectories based on \esref{GPE}{solitoneom} dashed teal line). (b,d) The two largest orbital populations $\lambda(t)$ from MCTDHB (dashed) and the two mode model \bref{twomode} (solid). For the latter we used a time-dependent soliton separation $d(t)$, which is inferred from the MCTDH peak densities. 
(e,g) Pre-collision atom number probabilities $\rho_n$ in the left soliton from the TMM at the times $t_a$ indicated by ($\star$) in (b,d), {the total number is fixed at $2000$, with on average $1000$ in each soliton}.
(f,h) The same after the collision, at times $t_b$ in (b,d). The figure uses dimensionless units as discussed in \aref{dimless_app}.}
\end{figure}

Let us consider collisions before fragmentation first. We see in \frefp{collisions_prior}{a,c} that quantum many-body theory and mean-field theory
agree on the character of collisions in this case. Most notably the initial relative phase controls whether interactions are attractive or repulsive. 
Note that the color shading indicated the total or mean atomic density from MCTDHB, which contains contributions from all orbitals present in \bref{mcfieldop} and depends on the quantum state. Collisions actually occur slightly earlier than the estimate $\sub{t}{coll}=|\sub{d}{ini}/(2\sub{v}{ini})|$, due the finite range of inter-soliton interactions.
The post-collision deviations of the trajectories visible in panel (a) is commented upon in \sref{sec_number_change}.

In addition to the atom density and hence trajectories, MCTDHB also provides us with the time-evolution of the eigenvalues of the OBDM $\lambda_\pm$, shown in panels (b,d). We compare  these with the $\lambda_\pm$ obtained from the TMM discussed in \sref{tmm}, with trajectories $d(t)$ adjusted to those in MCTDHB. It is apparent that collisions indeed occur prior to fragmentation, and the two models yield similar OBDM eigenvalues. The TMM now additionally allows us to inspect the atom number distribution in the left soliton $\rho_n = \sum_m |c_{nm}|^2$, see \sref{phasediff}.

We show this distribution in \frefp{collisions_prior}{e-h} at the times indicated by ($\star$) in panels (b,d), which are chosen just before and just after the collision. Outside of the time-window $[t_a, t_b]$, the number distribution is essentially conserved. The early snapshots at $t_a$ in panels (e,g) thus simply show the Gaussian $\rho_n$ for the initial relative coherent state $\ket{\sub{N}{tot},\pm}$. However, during closest approach, near $\sub{t}{coll}$ atom transfer terms containing the operator $\hat{b}^\dagger \hat{a} + \hat{a}^\dagger \hat{b}$ become large
in \bref{twomodeHamil} (terms $\sim J,\bar{J}$). Atoms can thus make transfers from one soliton to the other. This intermittent Bosonic-Josephon-Junction (BJJ) \cite{Albiez:oberthaler:BJJ},  causes a widening of the number distribution for the initial phase $\varphi=0$, see panel (f). This wider number distribution then accelerates the phase diffusion effect discussed in \sref{phasediff} and causes subsequent fragmentation already around $\sub{t}{frag}=15$, where without the collisions it would have only happened at $\sub{t}{frag}\approx60$. Note however, that the TMM results for the $\varphi=0$ may not be reliable, since exactly at the moment of collision the two chosen modes cease to be orthogonal. However, this is not a problem shared by MCTDH, which qualitatively agrees on an increase of the degree of fragmentation following the collision, albeit less severe. We thus conclude that attractive collisions will cause earlier subsequent fragmentation. 

In contrast, the number distribution is not significantly widened in the repulsively interacting case in panel (h), due to much weaker tunnelling. Note that this is not alone due to the minimal separation being larger in the repulsive case than in the attractive case: interaction terms become almost as large as in the $\varphi=0$ case anyway. Thus the $\varphi=\pi$ phase relation must be less conducive to atom transfer.

%%%%%%%%%%%%%%%%%%%%%%%%%%%%%%%%%%%%%%%%%%%%%%%%%%%%%%%%%%%%%%%%%%%%%%%%%%%%%%%%
\subsection{After fragmentation}
\label{after_frag}
%%%%%%%%%%%%%%%%%%%%%%%%%%%%%%%%%%%%%%%%%%%%%%%%%%%%%%%%%%%%%%%%%%%%%%%%%%%%%%%%

\begin{figure}[htb]
\includegraphics[width=0.99\columnwidth]{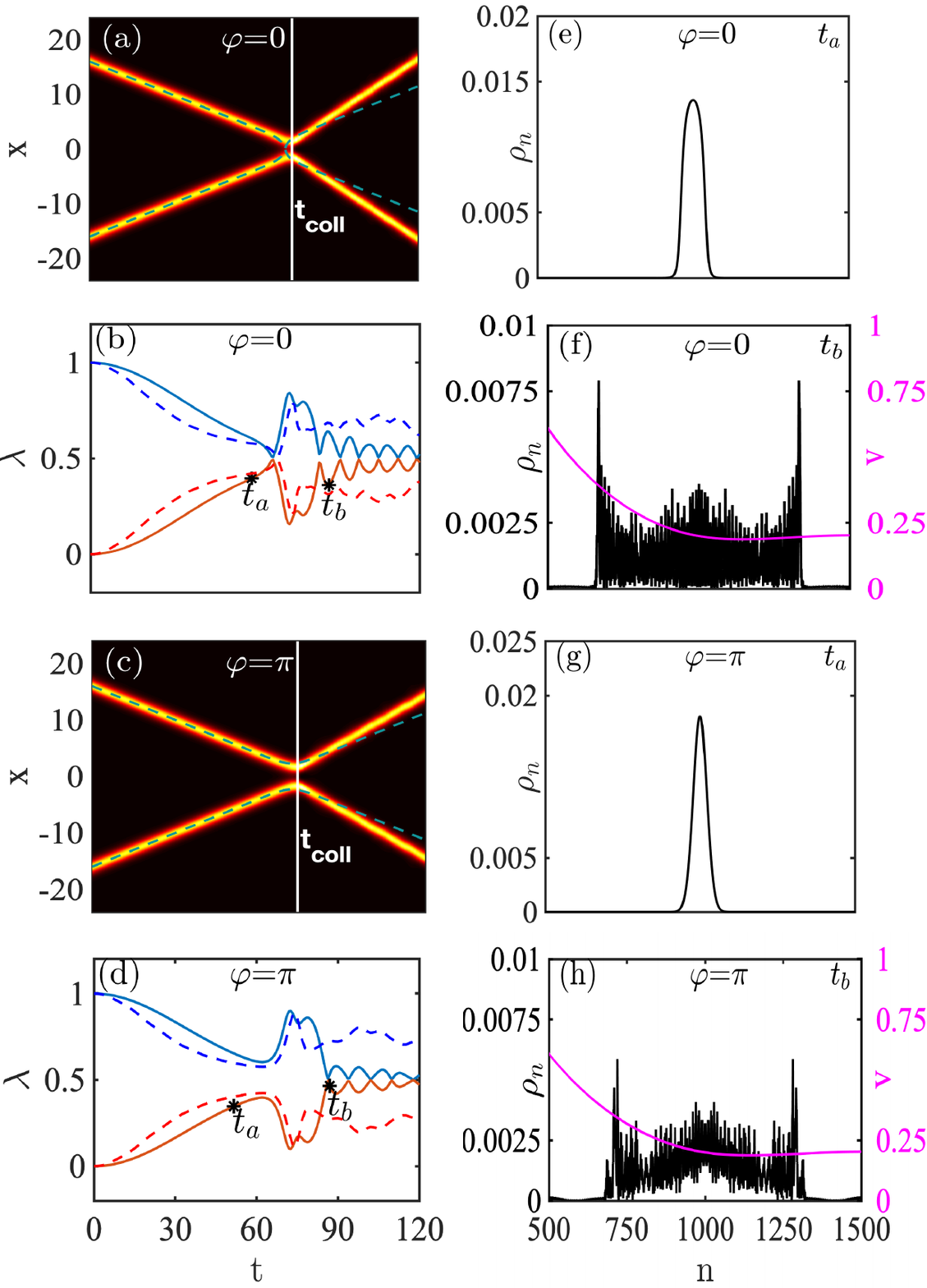}
\caption{\label{collisions_post} Collision and coherence dynamics in controlled soliton collisions \emph{after} fragmentation, $\sub{t}{coll}>\sub{t}{frag}$. Panel layout and curves as in \fref{collisions_prior}. Magenta lines and axes in panels (f,h) additionally show the dependence of post-collision velocity $v=p_+/m$ on atom number per soliton, see \sref{sec_momentum_balance} and \aref{velocity_of_a}.}
\end{figure}

We now move to collisions after fragmentation, $\sub{t}{coll}>\sub{t}{frag}$. In that case almost no initial phase-dependence of collision kinematics remains in the mean atomic density provided by MCTDHB, see \frefp{collisions_post}{a,c}. Mean collision trajectories always seem to have repulsive character, fairly regardless of the \emph{initial} relative phase between the solitons. It has been shown in \cite{sakman_singleshot_nphys}, however,  that an ``always repulsive'' appearance of the MCTDHB total atomic density may be misleading. The authors of \rref{sakman_singleshot_nphys} include all available information on the many-body wave function to predict the atom density for single realisations of the many-atom probability distribution for fragmented collisions, instead of the mean density that one would obtain by averaging many such realisations. Following these many-body collisions in time, one identifies collision trajectories akin to mean-field ones, with seeming random phases from realisation to realisation, including some attractive collisions.

We see the same behaviour in TWA collisions from a fully fragmented state. Also there, single trajectories are a random mix, exhibiting collisions that match the mean field picture for all relative phase angles $\varphi\in[0,2\pi)$ between solitons. A majority of these collisions have a repulsive ``appearance", thus a density average over all such trajectories yields a repulsive mean trajectory.  Features of both simulation techniques are consistent with the picture in \sref{phasediff}: After complete phase diffusion, all relative phases between the two solitons are part of the two soliton quantum state, so an individual collision may with some probability appear repulsive and with another attractive.

%%%%%%%%%%%%%%%%%%%%%%%%%%%%%%%%%%%%%%%%%%%%%%%%%%%%%%%%%%%%%%%%%%%%%%%%%%%%%%%%
\subsection{Collisions with number change}
\label{sec_number_change}
%%%%%%%%%%%%%%%%%%%%%%%%%%%%%%%%%%%%%%%%%%%%%%%%%%%%%%%%%%%%%%%%%%%%%%%%%%%%%%%%

\begin{figure}
\includegraphics[width=8.5cm]{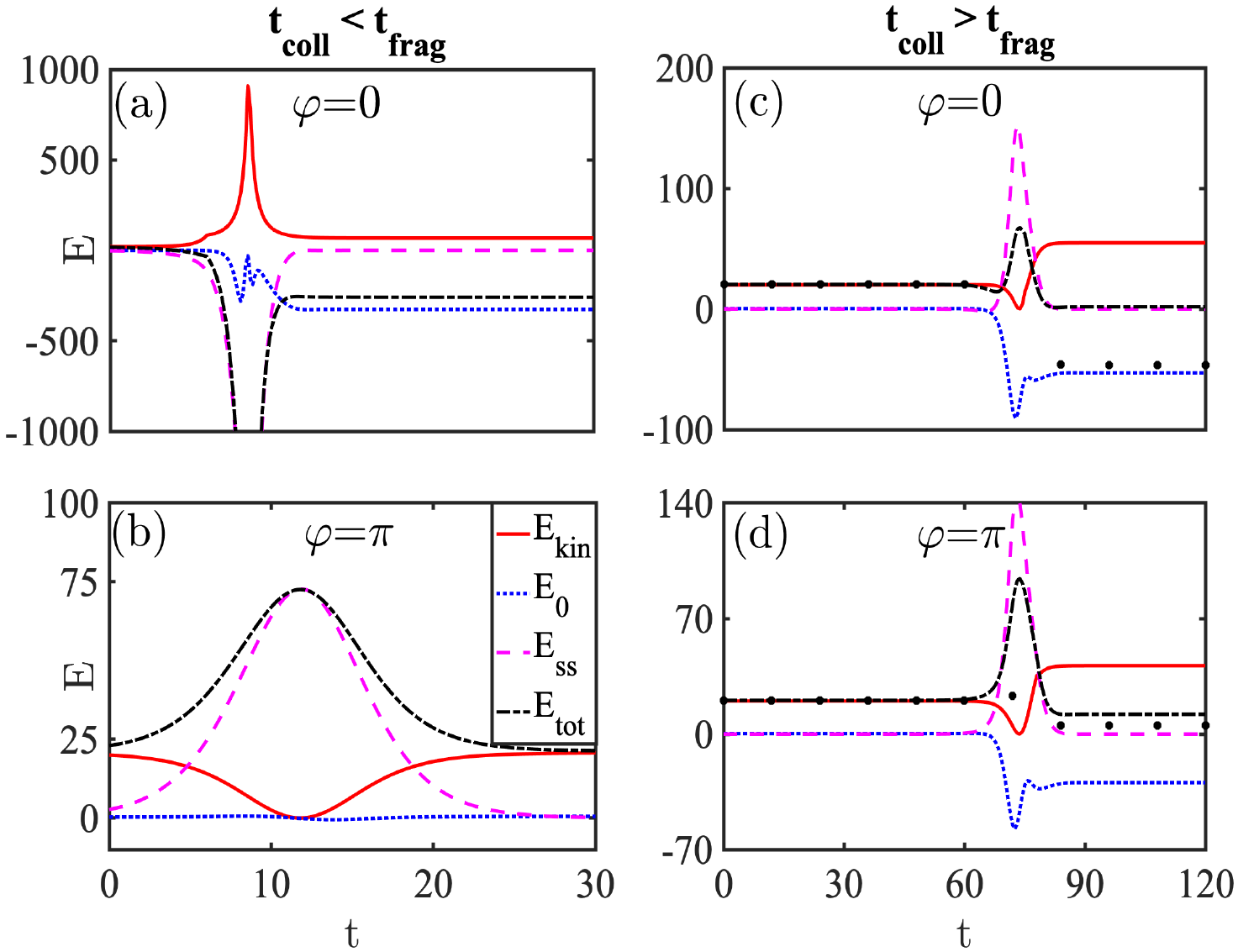}
\caption{\label{TMM_energies}  Conversion of mean interaction energy into kinetic energy during a soliton collision, using TMM with soliton distance $d(t)$ taken from MCTDHB. We show the total energy $\sub{E}{tot}$, \eref{EWtotTMM} (black dot-dashed), the kinetic energy, \eref{EkinTMM} (solid red) $\sub{E}{kin}$, the soliton self-interaction energy (dotted blue) $\sub{E}{0}$, \eref{EchiTMM} and inter-soliton interaction energy (dashed magenta) $\sub{E}{ss}$, defined in the text. (a,b) collision before fragmentation with (a) $\varphi=0$, (b) $\varphi=\pi$. (c,d) collision after fragmentation with (c) $\varphi=0$, (d) $\varphi=\pi$. {In (c,d), black $(\bullet)$ show $\sub{E}{tot}$ if we do not include an increase in kinetic energy.}}
\end{figure}

Besides the apparent indifference of mean collisions to the initial inter-soliton phase, a second prominent feature of \fref{collisions_post} is that MCTDHB predicts collisions 
to be \emph{super-elastic}, with solitons gaining kinetic energy in the collision, while total energy is conserved. This feature was also visible in panel (a) of \fref{collisions_prior}.
To understand possible physical reasons for this, we firstly multiplied the rhs of the soliton kinetic equation \bref{solitoneom} used for the TMM with a scale factor $f(t)$, phenomenologically adjusted to give trajectories in agreement with MCTDHB, i.e.~speeding up in the collision. We can then get a first idea of the source of additional kinetic energy, by inspecting the different contributions to the total energy 
\begin{align}
\sub{E}{tot}&=\expec{\hat{H}}+ \sub{E}{kin},
\label{EWtotTMM}
\end{align}
within the corresponding TMM in \fref{TMM_energies}.  

We can obtain $\expec{\hat{H}}$ from \bref{twomodeHamil}, while the joint kinetic energy of both solitons, each with velocity $\dot{d}(t)/2$, is
\begin{align}
\sub{E}{kin} = 2\times \frac{1}{2}m \sub{N}{sol}\left(\frac{\dot{d}(t)}{2}\right)^2.
\label{EkinTMM}
\end{align}
We then further split $\expec{\hat{H}}$ into a contribution internal to the solitons
\begin{align}
\sub{E}{0}= \langle \omega( \hat{a}^\dagger \hat{a} + \hat{b}^\dagger \hat{b} ) +  \frac{\chi}{2} ( \hat{a}^\dagger \hat{a}^\dagger \hat{a} \hat{a} + \hat{b}^\dagger \hat{b}^\dagger \hat{b} \hat{b}) \rangle
\label{EchiTMM}
\end{align}
and a soliton-soliton interaction energy $\sub{E}{ss}$ (all other terms of \bref{twomodeHamil}). For large separations $d(t)$, we must have $\sub{E}{ss}\rightarrow0$. 

We plot all energy contributions in \fref{TMM_energies}, setting the initial value of $\sub{E}{0}$ to zero, to ease the comparison of temporal changes.
We see in panels (c,d) of \fref{TMM_energies}, that there the drop in internal soliton energy, $\sub{E}{0}$, provides the extra kinetic energy found after collisions. We identify the atomic transfer between the solitons discussed earlier as cause for this, due to interactions of the form $\hat{J}(d)(\hat{b}^\dagger \hat{a} + \hat{a}^\dagger \hat{b})$. Here the coefficient $\hat{J}(d)$ depends on the overlap of the left and right soliton modes, and is relevant only briefly around the moment of collision. As shown in \frefp{collisions_post}{b,d}, the term causes significant restoration of phase coherence, with an accompanying widening of the atom number distribution $\rho_n$ in each soliton \frefp{collisions_post}{f,h}. This is in accordance with number and phase being conjugate variables. 

Since the internal energy per soliton $\sub{E}{0}\approx\chi \sum_n \rho_n n^2$ is negative and non-linearly dependent on atom number, a increase of the atom number uncertainty and thus widening of the distribution $\rho_n$ causes an internal energy drop $\Delta\sub{E}{0}\approx\chi \sum_n \Delta \rho_n n^2$, where $\Delta \rho_n$ is the difference between the number distributions before and after the collision. {We find that this can quantitatively explain the gain in kinetic energy as shown in \fref{TMM_energies}(c,d), up to a minor mismatch.

This minor mismatch is not surprising since we combine information from two independent methods (MCTDHB and TMM) that are not expected 
to be consistent in this combination. The point we stress is, that if solitons gain kinetic energy, the drop in internal energy is partially balanced in contrast to elastically colliding solitons, as seen by comparing the two black lines in \frefp{TMM_energies}{c,d}.}
The explanation works less well for panel (a). However, in that case the TMM is expected to break down at the moment of the collision, since the two modes become identical in that effectively attractive case. 

%%%%%%%%%%%%%%%%%%%%%%%%%%%%%%%%%%%%%%%%%%%%%%%%%%%%%%%%%%%%%%%%%%%%%%%%%%%%%%%%
\subsection{Momentum balance in collisions with number change}
\label{sec_momentum_balance}
%%%%%%%%%%%%%%%%%%%%%%%%%%%%%%%%%%%%%%%%%%%%%%%%%%%%%%%%%%%%%%%%%%%%%%%%%%%%%%%%

\begin{figure}
\includegraphics[width=0.69\columnwidth]{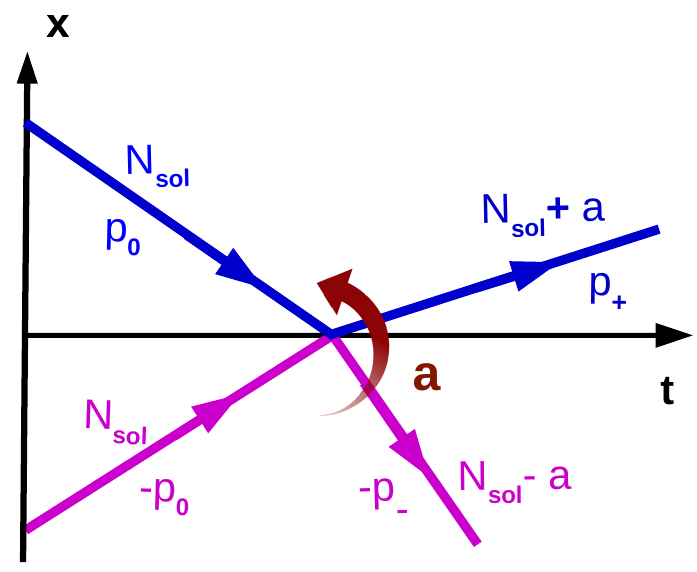}
\caption{\label{momentum_balance}
Momenta involved in a collision with exchange of atoms. The incoming solitons on the left of the graph contain $\sub{N}{sol}$ atoms, with momentum $\pm p_0$ respectively.
If $a$ atoms transfer from one soliton to the other, the larger one must move slower after the collision to conserve the total momentum. This is the case in a single realisation of the quantum many body superposition state.}
\end{figure}

However now that we have linked the increase in post-collision mean kinetic energy of solitons with atoms transferring from one soliton to the other, 
we must consider the implications of this picture, when taking into account momentum conservation. To this end we refer to \fref{momentum_balance}. For simplicity of the following argument, assume an equal number of atoms, $\sub{N}{sol}$, are contained in the two incoming solitons with momenta $p_{0}$ and -$p_{0}$ per atom sketched in \fref{momentum_balance}, thus the initial total net momentum is zero.

At the moment of collision, due to close proximity of solitons, atom transfer from one to the other is likely. Let us assume $a$ atoms are transferred from the left to the right soliton. If we denote the outgoing momenta per atom by $p_+$ and -$p_-$, conservation of momentum gives:
\begin{align}
(N_{sol}+a) p_{+} - (N_{sol}-a )p_{-} = 0,
\label{momcons}
\end{align}
which for $a>0$ already clearly requires $|p_{-}|>|p_{+} |$ as sketched in the figure.

An additional constraint arises from energy conservation
\begin{align}
\sub{N}{sol} \frac{p_0^2}{m} + \chi \sub{N}{sol}^2 &=(\sub{N}{sol}+a)\frac{p_+^2}{2m} + \chi\frac{(\sub{N}{sol}+a)^2}{2}\CR
&+(\sub{N}{sol}-a)\frac{p_-^2}{2m} + \chi\frac{(\sub{N}{sol}-a)^2}{2}.
\label{encons}
\end{align}
The equations \bref{momcons} and \bref{encons}  can be solved to yield momenta of atoms in outgoing solitons $p_\pm$ as a function of their initial constituent number $\sub{N}{sol}$, the number of atoms transferred in the collision $a$, initial momentum per atom $p_0$ and Hamiltonian parameters $m$, $\chi$. We find
\begin{align}
p_{+}& = \pm \frac{\sqrt {a-N_{sol}} \sqrt {a^{2} m \chi - p_{0}^{2} N_{sol} }}{\sqrt{a N_{sol}+ N_{sol}^{2}}}.
\end{align}
The resultant velocity $v(n)=p_+/m$ as a function of soliton constituent number $n=\sub{N}{sol}+a$ is shown as magenta lines in \frefp{collisions_post}{f,h}.

Importantly, $v(n)$ is not symmetric about $\sub{N}{sol}$ and \emph{non-linear}, such that if we calculate the mean outgoing kinetic energy $\sub{\bar{E}}{kin}$ as average of kinetic energies $ \sub{E}{kin}[v(n)]=m v(n)^2/2$ over the distribution of atom numbers in the soliton $\sub{\bar{E}}{kin}=\sum_n \rho_n \sub{E}{kin}[v(n)]$, the result can be \emph{larger} than the ingoing kinetic energy, and agrees quite closely with the MCTDHB proposal. On average, we can view this kinetic energy gain as fuelled by a drop in the internal soliton energy due to a widening of $\rho_n$. Of course, note, that the average atom transfer $\overline{a}$ must be zero by symmetry, thus if transfer of $a$ atoms occurs
with some probability, the same is true for $-a$. 

In the discussion of \sref{sec_momentum_balance} so far, we have neglected the initial atom number unncertainty required to implement a defined inter-soliton phase. Based on \frefp{collisions_post}{e-h}, these are small compared to fluctuations generated through atom transfer. 
 
We thus propose the following: in the super-elastic cases, the MCTDHB method provides a variationally optimised approximation within its two-mode constraint, to describe an entangling quantum many body collision beyond its reach: According to the arguments above, we would conclude that the post-collision soliton state for the two solitons is mesoscopically entangled, with a superposition of solitons of different constituent numbers located at different \emph{positions}, since they have moved with different velocities.  
Schematically we can write this state as
\begin{align}
\ket{\sub{\Psi}{pc}}&=\sum_{n_s} c_{n_s} \ket{n_s,v(n_s)}_{L} \CR
&\otimes \ket{2\sub{N}{sol}-n_s,v(2\sub{N}{sol}-n_s)}_{R},
\label{postcollstate}
\end{align}           
where $\ket{n,v}$ indicate the constituent number $n$ and velocity $v$ (hence also position) of the left and right soliton separately and $c_{n_s}$ are complex coefficients.
 
Of course, this also implies that the TMM and MCTDHB, which have provided this picture, cannot be valid for times much after the collision since their restriction to a single spatial mode or orbital per soliton precludes the description of an entangled state of position such as \bref{postcollstate}. For that one orbital per soliton and per contributing velocity class would be required. However the two physical causes of this final state, phase diffusion before collisions and atom transfer at the moment of collision both occur during the time in which the models are expected to be valid for the effectively repulsive collisions.  We thus expect our conclusions to persist qualitatively, unless some essential conservation law was broken by the approximations in the methods discussed in \sref{bmft}. One such conservation law would be provided by microscopic momentum conservation in a strictly 1D setting, but not in 3D as discussed in the next section.

The state \bref{postcollstate} is motivated by the processes discussed with evidence provided by the effective models used here. Once allowing significant non mean-field effects, other possibilities that those models could not have hinted at are for example the emission of atoms as radiation from the solitons \cite{Kevrekidis_radiation_PhysRevA}.

The schematic \bref{postcollstate} constitutes the many-body generalisation of semi-classical results \cite{Lewenstein_phasekin_entangle} and is also reminiscent of the collision induced two species Bell states proposed in \cite{Gertjerenken_cat_coll_PRL} and entanglement generation involving dark \cite{Mishmash_entangleddarksol_PRL} or dark-bright solitons \cite{Katsimiga_darkbright_NJP2017}.

%%%%%%%%%%%%%%%%%%%%%%%%%%%%%%%%%%%%%%%%%%%%%
\section{Integrability breaking} 
\label{integrability}
%%%%%%%%%%%%%%%%%%%%%%%%%%%%%%%%%%%%%%%%%%%%%

Let us now discuss the connection of the previous section with the absence of many-body integrability of the underlying model.
In a more extremely 1D scenario, where \emph{individual atomic collisions} can also be restricted to the single dimension $x$, a 1D variant of the Hamiltonian \bref{Hamiltonian} would become that of the Lieb-Liniger-MacGuire (LL) model \cite{McGuire_exactlysolvable_JMP,LL_model_PR}.
This model is integrable with an exact many-body solution, which contains the feature that the set of \emph{individual} atomic momenta is conserved \cite{McGuire_exactlysolvable_JMP,Holdaway_entanglesol,zhu_manybody_liebliniger_ChinPhysB}. Intuitively, in a setting such as \fref{momentum_balance}, the atoms within a soliton co-propagate in the incoming state, all with individual momenta either $p_0$, if they are in the left soliton, or $-p_0$ if they are in the right one. The binary delta function interaction potential $\delta(x_1-x_2)$ between atoms $1$, $2$, then can change the momentum of the atom pair only as in ($p_0$, $-p_0$) $\rightarrow$ ($-p_0$, $p_0$), i.e.~a completely elastic momentum flip collisions. This would preclude atom transfer processes as described in \sref{sec_momentum_balance}, since the momenta would always remain distributed at $N/2$ atoms with $p_0$ and $N/2$ atoms with $-p_0$.

While this \emph{collisionally} 1D regime has attracted considerable experimental attention \cite{kinoshita_cradle,hofferberth_quantum_thermal,hofferberth_1dcoherence}, it is not at all reached in any of the soliton experiments discussed in this article, see \sref{sec_experiments}. There, atoms are much more weakly confined transversely and collisions are thus 3D, significantly breaking integrability \cite{Mazets_breakinteg_PhysRevLett}. Also the presence of a harmonic trap in the longitudinal direction contributes to integrability breaking \cite{Holdaway_entanglesol}.
In such a scenario, methods in which many-body integrability is implicitly broken, which applies to all those presented in our \sref{bmft}, will provide \emph{qualitatively more physical} results than an artificially integrable method would. For example consider the free expansion of a repulsively interacting quasi-1D condensate in a wave guide as in the experiment \cite{Bongs_waveguide_PhysRevA}: Here the initial interaction energy is converted into kinetic energy by collisions, causing the momentum distribution to widen dynamically. This effect is not captured by the LL model, but is \emph{quantitatively} captured by the quasi-1D GPE \bref{GPE}. In the same sense we believe our methods paint a more physical picture of quasi-1D solitary wave collisions than the LL model would, which indeed does not contain atom transfer \cite{lai_quantsol_II}. Other work dealing with broken integrability in the context of soliton collisions also reported signs of atom transfer \cite{Khaykovich_quinticsol_PhysRevA,Holdaway_entanglesol}.

The connection of atom exchange during soliton collisions and (non-)integrability warrants further studies. Experiments could vary the degree of transverse confinement to enforce a transition between the regimes, while theory can explicitly introduce integrability breaking terms as in \cite{Holdaway_entanglesol} or \cite{Mazets_breakinteg_PhysRevLett} to the models discussed here. Besides these possibilities, the solitary wave collision scenario represents a surprisingly daunting scenario for theory: It would be desirable to employ a full fledged first principles simulation would have to deal with three spatial dimensions, mesoscopic entanglement and thermal noise (see \sref{disc_of_exp}) all at once. We are not aware of a formalism that is capable of all these at present. Nonetheless we have confidence in our conclusions, since they are based on two robust features of the underlying physics: (i) phase diffusion that is present in any condensate with number fluctuations and (ii) Josephson type tunnelling that would be present in any well defined two-mode system with contact between the modes. Both occur at times \emph{before} our approximation methods cease to be valid.

Finally, the discussion in this section implies that our methods cannot quantitatively predict the amplitude generation of new momenta as in \fref{sec_momentum_balance}, since this process must rely on many-body integrability breaking through the approximations leading to MCTDHB or the two mode model, the nature of which warrants further studies. For a quantitative prediction, physical integrability breaking terms \cite{Holdaway_entanglesol,Mazets_breakinteg_PhysRevLett} will be added in the future. Turning the argument around, our results also present clear evidence that the MCTDHB in one spatial dimension does not preserve the many-body integrability of the LL model when applied to it.

%%%%%%%%%%%%%%%%%%%%%%%%%%%%%%%%%%%%%%%%%%%%%
\section{Velocity dependence of atom transfer} 
\label{sec_velocity_dependence}
%%%%%%%%%%%%%%%%%%%%%%%%%%%%%%%%%%%%%%%%%%%%%
%
In \sref{sec_number_change} we had discussed that atom-transfer during a collision of solitons can lead to dramatic consequences for the final many-body quantum state. In the framework of the two-mode model \bref{twomode}, this transfer is due to non-adiabatic effects from the temporal change of the Hamiltonian $\hat{H}$ in \bref{twomodeHamil}. These changes arise from the coefficients $J$, $\bar{J}$, $\bar{D}$ that depend on soliton separation $d(t)$. 

\begin{figure}
\includegraphics[width=0.99\columnwidth]{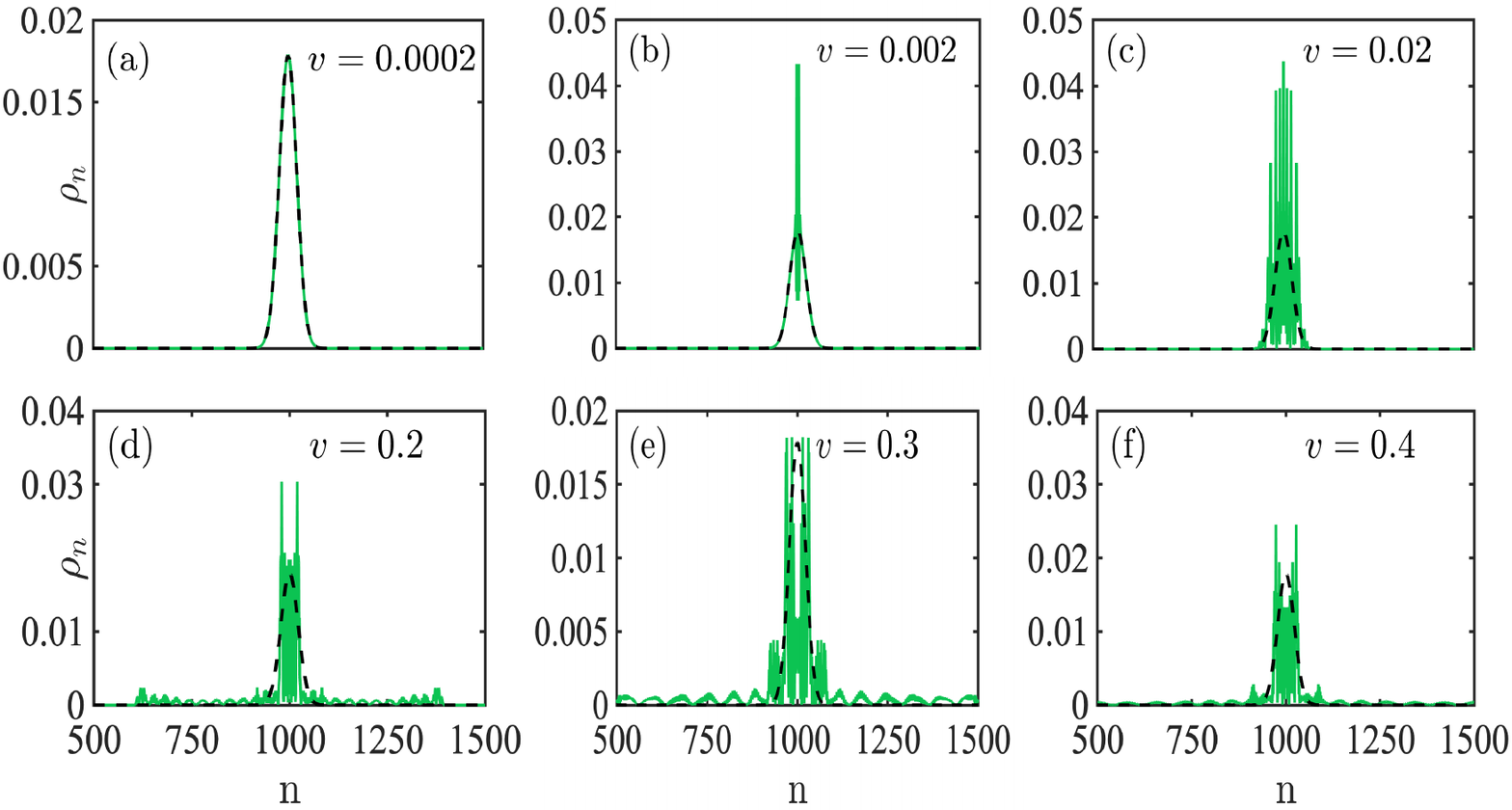}
\caption{\label{velocity_dependence}Atom number distribution $\rho_n$ in a single soliton from the TMM before the collision (black-dashed) and after (green solid). The scenario is the same as in \fref{collisions_post}, except that we vary the velocity of each soliton $v$ as indicated in the panels. All cases are for repulsive initial phases $\varphi=\pi$ and brought to collision at $\sub{t}{coll}=80$ or later (after fragmentation) by adjusting $\sub{d}{ini}$.}
\end{figure}

We thus expect the number distribution in soliton collisions to significantly depend on the collision velocity. However for very slow collisions, we expect the two-mode model quantum state to adiabatically follow the changes in parameters, and thus return to the initial state after the collisions, which is what is seen in \frefp{velocity_dependence}{a}. In the velocity range in between, \fref{velocity_dependence} show that the number transfer depends non-trivially on the collision velocity. 

In principle we could expect a second regimes without a change in the number distribution: For very fast collisions the quantum many-body state should stay unchanged, since a very short collision yields a sudden, impulsive change in Hamiltonian parameters with equally sudden return to the initial Hamiltonian. However we find that for collisions fast enough for that, there is sufficient kinetic energy for solitons to overcome their interactions even in the repulsive case. Since left and right soliton modes thus overlap at the collisions, results will be unreliable.

For repulsive collisions before fragmentation, we find no changes in the number distribution regardless of velocity.

%%%%%%%%%%%%%%%%%%%%%%%%%%%%%%%%%%%%%%%%%%%%%
\section{Soliton decoherence at non-zero temperature and discussion of experiments} 
\label{disc_of_exp}
%%%%%%%%%%%%%%%%%%%%%%%%%%%%%%%%%%%%%%%%%%%%%

We will now discuss how the predictions of the other sections are consistent with existing experiments on soliton trains and their interactions and can further answer a variety of hitherto open questions.

Our analytical model \bref{eigenvalues} predicts a fragmentation time of $\sub{t}{frag}=877$ ms for the TSE \cite{Nguyen_solcoll_controlled} assuming $T\approx0$ and $\sub{N}{sol}=28000$, $a_s=-0.57a_0$, $\omega_{\perp}=(2\pi)254$ Hz. This is substantially beyond the experimentally covered range of collision times $\sub{t}{coll}<30\dots320$ ms. However it is comparable with the initial preparation time $\sub{t}{prep}$, which exceeds $750$ ms. It takes that long in the experiment, to adiabatically split the condensate and then adjust the interactions strength from its initial repulsive value to the final attractive one. Throughout all this time, phase-diffusion will already be active. It is thus likely that TSE collisions already \emph{begin} in a phase diffused and fragmented state. This is consistent with the experimental observation that collisions are indicative of all phases in $[0,2\pi)$, see our discussion at the beginning of \sref{after_frag}. Once in-situ observation has collapsed a certain soliton pair onto a specific relative phase, the subsequent time is too short for re-fragmentation and thus further collisions are consistent with that initially chosen mean-field relative phase. 

\begin{figure}[htb]
\includegraphics[width=0.99\columnwidth]{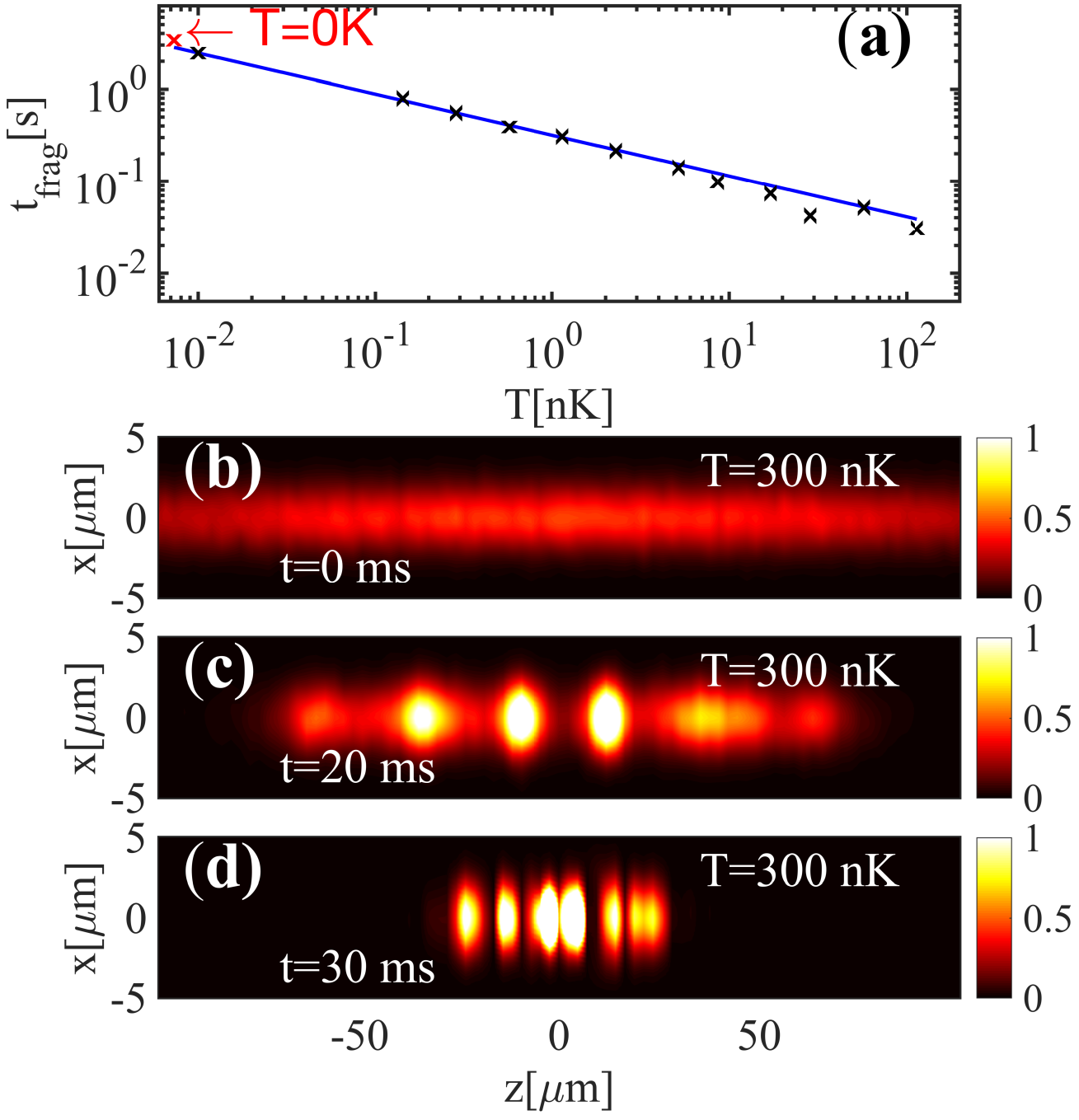}
\caption{\label{MSExperiments} (a) Fragmentation time as a function of temperature $T>0$ from TWA, for soliton parameters matching the multi-soliton-experiment \cite{Nguyen_modulinst}. The solid line is the fit $\sub{t}{frag}=0.32$s $(T/\mbox{nK})^{-0.44}$. (b,c) Normalized column-density from single TWA trajectory with thermal noise matching $\sub{T}{eff}=300$ nK, see also supplemental movies (black-zero, bright-high). Lower effective temperatures would not reproduce the initial fluctuations visible in the experiment, (b), nor the time of first formation of solitons in (c).
Panel (d) demonstrates the complete soliton train at later time. An initial uncondensed component with effective temperature $\sub{T}{eff}=300$nK would cause fragmentation on a $10$ ms time scale, according to panel (a). }
\end{figure}

To investigate the onset of fragmentation and its dynamics, $\sub{t}{frag}$ should be reduced. Larger solitons or stronger interactions could be problematic due to losses, but one can employ higher temperatures or noise, as we show now.  Finite temperature condensates can straightforwardly be modelled using the TWA \cite{blair:review}. Returning to the scenario of two non-colliding solitons identical to the one in \fref{fragmentation_allmethods}, we show the temperature dependence of the fragmentation time-scale in \frefp{MSExperiments}{a}.  The data is fit by $\sub{t}{frag}\sim T^{-0.44}$. The additional spread of inter-soliton phases due to the interaction with hotter uncondensed atoms thus significantly accelerates fragmentation. 
It is useful that $\sub{t}{frag}$ spans the full range, from longer than most experiments ($\sim1$s), down to shorter than many ($\sim50$ ms), within the relevant temperature range from a few nK to typical condensation temperatures of a few $100$ nK. This opens a convenient window on the intricate many body dynamics described in earlier sections, while still permitting to avoid fragmentation for interferometric applications.

In the light of accelerated fragmentation due to thermal atoms, let us now also revisit the MSE \cite{Nguyen_modulinst}. We performed a 3D simulation of that experiment, using a single-trajectory of TWA at finite temperature. Column densities as shown in \fref{MSExperiments}, that correspond to an image of the atomic cloud taken from the side, should roughly agree with those in \cite{Nguyen_modulinst}, regarding characteristic features like amplitude of fluctuations or formation time of solitons. This is only possible by assuming relatively high initial effective temperatures $\sub{T}{eff}\gtrsim300$ nK. Referring to \frefp{MSExperiments}{a}, for which $\sub{N}{sol}$ and $U_0$ are matching this experiment, we then read off an expected fragmentation time of the order of $\sub{t}{frag}={\cal O}$($10$ ms), compared to  $\sub{t}{frag}\approx2$ s  at $T=0$, based on $\sub{N}{sol}=40000$, $a_s=-0.18a_0$,  $\omega_{\perp}=(2\pi)346$ Hz. Only under these conditions, the entire soliton-train can fragment before the moment of first collisions, about $\sub{t}{coll}=15$ ms after soliton formation and $\sub{t}{coll}+\sub{t}{form}=25$ ms after experiment initiation. Subsequent collisions would then be expected to have predominantly repulsive character as experimentally observed.

 A hot initial condensate is an even more appropriate starting point for the earlier experiments that reported mainly repulsion in soliton trains \cite{strecker:brighsol,jila:solitons}, which first went through collapse instabilities causing substantial non-equilibrium heating \cite{jila:nova,wuester:nova,wuester:nova2,holland:burst}. In contrast, accelerated fragmentation due to environmental noise does not occur during collisions in the TSE, since soliton creation there initially follows a slow adiabatic procedure, with substantially less heating than during a collapse or instability.

Another prediction of this article that contributes to an overall picture of predominantly repulsive collisions due to quantum effects, is the acceleration of fragmentation if there are attractive collisions (possibly initially and rare), as discussed in \sref{before_frag}.  While most of the experiments discussed would not have had the sensitivity to detect the super-elastic effects predicted in \sref{sec_number_change}, these should play a role in the TSE setting \cite{Nguyen_solcoll_controlled}, and could possibly be observed with minor improvements of the sensitivity there.

Finally, the dynamic choice of a fixed relative phase in the TSE would be related to measurement induced collapse of the many-body wave function, according to the picture here. The possibility to  continuously and non-destructively infer soliton collisions properties in a setup such as \cite{Nguyen_solcoll_controlled} opens the door wide for explorations of the interplay between the highly entangling many-body collision dynamics predicted in earlier sections and continuous, controlled wave function collapse by measurements. 

%%%%%%%%%%%%%%%%%%%%%%%%%%%%%%%%%%%%%%%%%%%%%
\section{Comparison of methods} 
\label{com}
%%%%%%%%%%%%%%%%%%%%%%%%%%%%%%%%%%%%%%%%%%%%%

Even though TWA formally should be valid only closer to a mean-field situation, it agrees with MCTDHB on a large number of features in post-fragmentation soliton collisions: (i) The fact that these are a mixture of repulsive and attractive ones, with more repulsive ones, (ii) the qualitative shape of mean-density and (iii) the re-coherence features evident in \frefp{collisions_post}{f,h}.
We would like to place this observation in the context of the discussion in \cite{sakman_singleshot_nphys,dummond:twamctdh:comment,sakman:twamctdh:comment,alon:attractive:anharmonic,Cosme_com_motion}.

The present work demonstrates fruitful complementarity of all three methods employed: Thermal effects in \fref{MSExperiments} are naturally treated in the TWA. TWA however is troubled by controlled collisions as in \fref{collisions_post}, since it must also include random velocities and positions of the solitons. The latter yield an uncertain $\sub{t}{coll}$, blurring collisions when averaging. These fluctuations are inherent in the quantum dynamics of the centre of mass (CM) wave-function of solitons \cite{weiss_CMdiffusion,Cosme_com_motion}, but not included in MCTDHB with two orbitals. In contrast to \cite{Cosme_com_motion} we consider this a positive feature: 
 the absence of CM diffusion in MCTDHB simplifies studies of collisions. At the same time
agreement where possible between TWA and MCTDHB and consistency with our physical mechanisms makes us confident that TWA and MCTDHB have captured the essential many-body dynamics of phase-diffusion or fragmentation correctly. To pin-point the underlying basic physics on the other hand, reduction to the simple two-mode model has been most useful. 

%%%%%%%%%%%%%%%%%%%%%%%%%%%%%%%%
\section{Conclusions and Outlook} 
%%%%%%%%%%%%%%%%%%%%%%%%%%%%%%%%

We comprehensively consider two crucial beyond-mean-field features in soliton collisions. The first, phase-diffusion is clearly linked to the fragmentation of soliton trains reported in \cite{streltsov_frag}. Phase-diffusion occurs whenever the atom-number within one soliton is uncertain.
Since we cannot allocate a well-defined inter-soliton phase without allowing an uncertain atom-number due to their complementarity, it is thus unavoidable in principle that fragmentation eventually invalidates mean-field theory for soliton collisions. In practice the relevant time-scale, which we have evaluated analytically, can be fairly large at very low temperatures. 

We have further shown that the time-scale is shortened significantly through the presence of un-condensed atoms, whether these arise from non-zero temperature or non-equilibrium dynamics. Through this acceleration of fragmentation, beyond mean field effects can explain predominantly repulsive interactions of solitons in trains generated after some non-equilibrium instability dynamics \cite{jila:solitons,strecker:brighsol,Nguyen_modulinst}. Nonetheless, we still expect soliton collisions under more controlled conditions as in \cite{Nguyen_solcoll_controlled} to adhere to mean-field theory.

We have additionally {suggested} the generation of entanglement between atom-number and post-collisions position and momentum through soliton collisions involving integrability breaking features of the underlying 3D physics. {These features warrant further explorations, using either full-fledged 3D quantum field methods in a top-down approach, or by explicitly including physical processes that break many-body integrability in effective 1D models.}

%%%%%%%%%%%%%%%%%%%%%%%%%%%%%%%%
\acknowledgments
We gladly acknowledge the Max-Planck society for funding under the MPG-IISER partner group program, and interesting discussions with T.~Busch, S.~Cornish, M.~Davis, S. Gardiner, J.~Hope and C.~Weiss. ASR acknowledges the Department of Science and Technology (DST), New Delhi, India, for the INSPIRE fellowship IF160381. \\
%%%%%%%%%%%%%%%%%%%%%%%%%%%%%%%%
\appendix
%%%%%%%%%%%%%%%%%%%%%%%%%%%%%%%%
%%%%%%%%%%%%%%%%%%%%%%%%%%%%%%%%
%%%%%%%%%%%%%%%%%%%%%%%%%%%%%%%\

%%%%%%%%%%%%%%%%%%%%%%%%%%%%%%%%
\section{Dimensionless units}
\label{dimless_app}
%%%%%%%%%%%%%%%%%%%%%%%%%%%%%%%%
%
The 1D GPE \bref{GPE} can be written in a dimensionless form, by transforming wavefunction, space and time co-ordinates respectively as $\tilde\phi$ = $\phi\sqrt{L}$, $\tilde x$ = $\frac{x}{L}$ and $\tilde t$ = $\frac{t}{T}$, where tilded quantities are dimensionless. The scales are
$T$ = $\frac{m L^{2} }{\hbar}$ and $L$ = $\frac{2 \hbar^{2} }{m |U_{0}| N_{sol} } $, where the latter is chosen to yield a dimensionless soliton size $\tilde{\xi}=1$ for our most commonly used parameters. After un-tilding all variables except $\tilde{U}_0=\frac{T}{L\hbar}U_0$, the dimensionless GPE is then
\begin{align}
i \frac{\partial}{\partial t}\phi(x,t) =  \left[ -\frac{1}{2} \frac{\partial^2}{\partial x^2}  + \tilde{U}_0{|\phi(x,t)|}^2\right] \phi(x,t).
\label{GPEdimless}
\end{align}
%

%%%%%%%%%%%%%%%%%%%%%%%%%%%%%%%%
\section{Post-collision velocity after atom transfer}
\label{velocity_of_a} 
%%%%%%%%%%%%%%%%%%%%%%%%%%%%%%%%

After solving equations \eref{momcons} and \eref{encons}, the outgoing momenta as a function of $N_{sol}$, a, m and $\chi$ are as follows:
\begin{align}
p_{+}& =\frac{\sqrt {a-N_{sol}} \sqrt {a^{2} m \chi - p_{0}^{2} N_{sol} }}{\sqrt{a N_{sol}+ N_{sol}^{2}}}.
\end{align}
Thus the resultant velocity takes the form, $v(n)$ = $\frac{p_{+}}{m}$ which is a function of $N_{sol}$. These are shown with the magenta line in panels (f,h) in \fref{collisions_post}
%%%%%%%%%%%%%%%%%%%%%%%%%%%%%%%%
\providecommand{\noopsort}[1]{}\providecommand{\singleletter}[1]{#1}%

%%%%%%%%%%%%%%%%%%%%%%%%%%%%%%%%

\begin{thebibliography}{0}%
\makeatletter
\providecommand \@ifxundefined [1]{%
 \@ifx{#1\undefined}
}%
\providecommand \@ifnum [1]{%
 \ifnum #1\expandafter \@firstoftwo
 \else \expandafter \@secondoftwo
 \fi
}%
\providecommand \@ifx [1]{%
 \ifx #1\expandafter \@firstoftwo
 \else \expandafter \@secondoftwo
 \fi
}%
\providecommand \natexlab [1]{#1}%
\providecommand \enquote  [1]{``#1''}%
\providecommand \bibnamefont  [1]{#1}%
\providecommand \bibfnamefont [1]{#1}%
\providecommand \citenamefont [1]{#1}%
\providecommand \href@noop [0]{\@secondoftwo}%
\providecommand \href [0]{\begingroup \@sanitize@url \@href}%
\providecommand \@href[1]{\@@startlink{#1}\@@href}%
\providecommand \@@href[1]{\endgroup#1\@@endlink}%
\providecommand \@sanitize@url [0]{\catcode `\\12\catcode `\$12\catcode
  `\&12\catcode `\#12\catcode `\^12\catcode `\_12\catcode `\%12\relax}%
\providecommand \@@startlink[1]{}%
\providecommand \@@endlink[0]{}%
\providecommand \url  [0]{\begingroup\@sanitize@url \@url }%
\providecommand \@url [1]{\endgroup\@href {#1}{\urlprefix }}%
\providecommand \urlprefix  [0]{URL }%
\providecommand \Eprint [0]{\href }%
\providecommand \doibase [0]{https://doi.org/}%
\providecommand \selectlanguage [0]{\@gobble}%
\providecommand \bibinfo  [0]{\@secondoftwo}%
\providecommand \bibfield  [0]{\@secondoftwo}%
\providecommand \translation [1]{[#1]}%
\providecommand \BibitemOpen [0]{}%
\providecommand \bibitemStop [0]{}%
\providecommand \bibitemNoStop [0]{.\EOS\space}%
\providecommand \EOS [0]{\spacefactor3000\relax}%
\providecommand \BibitemShut  [1]{\csname bibitem#1\endcsname}%
\let\auto@bib@innerbib\@empty
%</preamble>
\end{thebibliography}%


\begin{thebibliography}{90}
\expandafter\ifx\csname natexlab\endcsname\relax\def\natexlab#1{#1}\fi
\expandafter\ifx\csname bibnamefont\endcsname\relax
  \def\bibnamefont#1{#1}\fi
\expandafter\ifx\csname bibfnamefont\endcsname\relax
  \def\bibfnamefont#1{#1}\fi
\expandafter\ifx\csname citenamefont\endcsname\relax
  \def\citenamefont#1{#1}\fi
\expandafter\ifx\csname url\endcsname\relax
  \def\url#1{\texttt{#1}}\fi
\expandafter\ifx\csname urlprefix\endcsname\relax\def\urlprefix{URL }\fi
\providecommand{\bibinfo}[2]{#2}
\providecommand{\eprint}[2][]{\url{#2}}

\bibitem[{\citenamefont{Pethick and Smith}(2002)}]{book:pethik}
\bibinfo{author}{\bibfnamefont{C.~J.} \bibnamefont{Pethick}} \bibnamefont{and}
  \bibinfo{author}{\bibfnamefont{H.}~\bibnamefont{Smith}},
  \emph{\bibinfo{title}{{Bose-Einstein} condensation in dilute gases}}
  (\bibinfo{publisher}{Cambridge University Press}, \bibinfo{year}{2002}).

\bibitem[{\citenamefont{Dalfovo et~al.}(1999)\citenamefont{Dalfovo, Giorgini,
  Pitaevskii, and Stringari}}]{stringari:review}
\bibinfo{author}{\bibfnamefont{F.}~\bibnamefont{Dalfovo}},
  \bibinfo{author}{\bibfnamefont{S.}~\bibnamefont{Giorgini}},
  \bibinfo{author}{\bibfnamefont{L.~P.} \bibnamefont{Pitaevskii}},
  \bibnamefont{and}
  \bibinfo{author}{\bibfnamefont{S.}~\bibnamefont{Stringari}},
  \bibinfo{journal}{Rev. Mod. Phys.} \textbf{\bibinfo{volume}{71}},
  \bibinfo{pages}{463} (\bibinfo{year}{1999}).

\bibitem[{\citenamefont{Kivshar and Agrawal}(2003)}]{book:solitons}
\bibinfo{author}{\bibfnamefont{Y.~S.} \bibnamefont{Kivshar}} \bibnamefont{and}
  \bibinfo{author}{\bibfnamefont{G.~P.} \bibnamefont{Agrawal}},
  \emph{\bibinfo{title}{Optical Solitons: From Fibers to Photonic Crystals}}
  (\bibinfo{publisher}{Academic, San Diego}, \bibinfo{year}{2003}).

\bibitem[{\citenamefont{Strecker et~al.}(2003)\citenamefont{Strecker,
  Partridge, Truscott, and Hulet}}]{li_rev}
\bibinfo{author}{\bibfnamefont{K.~E.} \bibnamefont{Strecker}},
  \bibinfo{author}{\bibfnamefont{G.~B.} \bibnamefont{Partridge}},
  \bibinfo{author}{\bibfnamefont{A.~G.} \bibnamefont{Truscott}},
  \bibnamefont{and} \bibinfo{author}{\bibfnamefont{R.~G.} \bibnamefont{Hulet}},
  \bibinfo{journal}{New Journal of Physics} \textbf{\bibinfo{volume}{5}},
  \bibinfo{pages}{73} (\bibinfo{year}{2003}).

\bibitem[{\citenamefont{Khaykovich et~al.}(2002)\citenamefont{Khaykovich,
  Schreck, Ferrari, Bourdel, Cubizolles, Carr, Castin, and
  Salomon}}]{khay:brighsol}
\bibinfo{author}{\bibfnamefont{L.}~\bibnamefont{Khaykovich}},
  \bibinfo{author}{\bibfnamefont{F.}~\bibnamefont{Schreck}},
  \bibinfo{author}{\bibfnamefont{G.}~\bibnamefont{Ferrari}},
  \bibinfo{author}{\bibfnamefont{T.}~\bibnamefont{Bourdel}},
  \bibinfo{author}{\bibfnamefont{J.}~\bibnamefont{Cubizolles}},
  \bibinfo{author}{\bibfnamefont{L.~D.} \bibnamefont{Carr}},
  \bibinfo{author}{\bibfnamefont{Y.}~\bibnamefont{Castin}}, \bibnamefont{and}
  \bibinfo{author}{\bibfnamefont{C.}~\bibnamefont{Salomon}},
  \bibinfo{journal}{Science} \textbf{\bibinfo{volume}{296}},
  \bibinfo{pages}{1290} (\bibinfo{year}{2002}), ISSN \bibinfo{issn}{0036-8075}.

\bibitem[{\citenamefont{Strecker et~al.}(2002)\citenamefont{Strecker,
  Partridge, Truscott, and Hulet}}]{strecker:brighsol}
\bibinfo{author}{\bibfnamefont{K.~E.} \bibnamefont{Strecker}},
  \bibinfo{author}{\bibfnamefont{G.~B.} \bibnamefont{Partridge}},
  \bibinfo{author}{\bibfnamefont{A.~G.} \bibnamefont{Truscott}},
  \bibnamefont{and} \bibinfo{author}{\bibfnamefont{R.~G.} \bibnamefont{Hulet}},
  \bibinfo{journal}{Nature} \textbf{\bibinfo{volume}{417}},
  \bibinfo{pages}{150} (\bibinfo{year}{2002}).

\bibitem[{\citenamefont{Eiermann et~al.}(2004)\citenamefont{Eiermann, Anker,
  Albiez, Taglieber, Treutlein, Marzlin, and Oberthaler}}]{gap_exp}
\bibinfo{author}{\bibfnamefont{B.}~\bibnamefont{Eiermann}},
  \bibinfo{author}{\bibfnamefont{T.}~\bibnamefont{Anker}},
  \bibinfo{author}{\bibfnamefont{M.}~\bibnamefont{Albiez}},
  \bibinfo{author}{\bibfnamefont{M.}~\bibnamefont{Taglieber}},
  \bibinfo{author}{\bibfnamefont{P.}~\bibnamefont{Treutlein}},
  \bibinfo{author}{\bibfnamefont{K.-P.} \bibnamefont{Marzlin}},
  \bibnamefont{and} \bibinfo{author}{\bibfnamefont{M.~K.}
  \bibnamefont{Oberthaler}}, \bibinfo{journal}{Phys. Rev. Lett.}
  \textbf{\bibinfo{volume}{92}}, \bibinfo{pages}{230401}
  (\bibinfo{year}{2004}).

\bibitem[{\citenamefont{Cornish et~al.}(2006)\citenamefont{Cornish, Thompson,
  and Wieman}}]{jila:solitons}
\bibinfo{author}{\bibfnamefont{S.~L.} \bibnamefont{Cornish}},
  \bibinfo{author}{\bibfnamefont{S.~T.} \bibnamefont{Thompson}},
  \bibnamefont{and} \bibinfo{author}{\bibfnamefont{C.~E.}
  \bibnamefont{Wieman}}, \bibinfo{journal}{Phys. Rev. Lett.}
  \textbf{\bibinfo{volume}{96}}, \bibinfo{pages}{170401}
  (\bibinfo{year}{2006}).

\bibitem[{\citenamefont{Nguyen et~al.}(2017)\citenamefont{Nguyen, Luo, and
  Hulet}}]{Nguyen_modulinst}
\bibinfo{author}{\bibfnamefont{J.~H.~V.} \bibnamefont{Nguyen}},
  \bibinfo{author}{\bibfnamefont{D.}~\bibnamefont{Luo}}, \bibnamefont{and}
  \bibinfo{author}{\bibfnamefont{R.~G.} \bibnamefont{Hulet}},
  \bibinfo{journal}{Science} \textbf{\bibinfo{volume}{356}},
  \bibinfo{pages}{422} (\bibinfo{year}{2017}).

\bibitem[{\citenamefont{Nguyen et~al.}(2014)\citenamefont{Nguyen, Dyke, Luo,
  Malomed, and Hulet}}]{Nguyen_solcoll_controlled}
\bibinfo{author}{\bibfnamefont{J.~H.~V.} \bibnamefont{Nguyen}},
  \bibinfo{author}{\bibfnamefont{P.}~\bibnamefont{Dyke}},
  \bibinfo{author}{\bibfnamefont{D.}~\bibnamefont{Luo}},
  \bibinfo{author}{\bibfnamefont{B.~A.} \bibnamefont{Malomed}},
  \bibnamefont{and} \bibinfo{author}{\bibfnamefont{R.~G.} \bibnamefont{Hulet}},
  \bibinfo{journal}{Nature Physics} \textbf{\bibinfo{volume}{10}},
  \bibinfo{pages}{918} (\bibinfo{year}{2014}).

\bibitem[{\citenamefont{Marchant et~al.}(2016)\citenamefont{Marchant, Billam,
  Yu, Rakonjac, Helm, Polo, Weiss, Gardiner, and Cornish}}]{Marchant_Quantrefl}
\bibinfo{author}{\bibfnamefont{A.~L.} \bibnamefont{Marchant}},
  \bibinfo{author}{\bibfnamefont{T.~P.} \bibnamefont{Billam}},
  \bibinfo{author}{\bibfnamefont{M.~M.~H.} \bibnamefont{Yu}},
  \bibinfo{author}{\bibfnamefont{A.}~\bibnamefont{Rakonjac}},
  \bibinfo{author}{\bibfnamefont{J.~L.} \bibnamefont{Helm}},
  \bibinfo{author}{\bibfnamefont{J.}~\bibnamefont{Polo}},
  \bibinfo{author}{\bibfnamefont{C.}~\bibnamefont{Weiss}},
  \bibinfo{author}{\bibfnamefont{S.~A.} \bibnamefont{Gardiner}},
  \bibnamefont{and} \bibinfo{author}{\bibfnamefont{S.~L.}
  \bibnamefont{Cornish}}, \bibinfo{journal}{Phys. Rev. A}
  \textbf{\bibinfo{volume}{93}}, \bibinfo{pages}{021604(R)}
  (\bibinfo{year}{2016}).

\bibitem[{\citenamefont{Marchant et~al.}(2013)\citenamefont{Marchant, Billam,
  Wiles, Yu, Gardiner, and Cornish}}]{Marchant_controlledform}
\bibinfo{author}{\bibfnamefont{A.~L.} \bibnamefont{Marchant}},
  \bibinfo{author}{\bibfnamefont{T.~P.} \bibnamefont{Billam}},
  \bibinfo{author}{\bibfnamefont{T.~P.} \bibnamefont{Wiles}},
  \bibinfo{author}{\bibfnamefont{M.~M.~H.} \bibnamefont{Yu}},
  \bibinfo{author}{\bibfnamefont{S.~A.} \bibnamefont{Gardiner}},
  \bibnamefont{and} \bibinfo{author}{\bibfnamefont{S.~L.}
  \bibnamefont{Cornish}}, \bibinfo{journal}{Nature Comm.}
  \textbf{\bibinfo{volume}{4}}, \bibinfo{pages}{1865} (\bibinfo{year}{2013}).

\bibitem[{\citenamefont{Medley et~al.}(2014)\citenamefont{Medley, Minar, Cizek,
  Berryrieser, and Kasevich}}]{Medley_evapsoliton}
\bibinfo{author}{\bibfnamefont{P.}~\bibnamefont{Medley}},
  \bibinfo{author}{\bibfnamefont{M.~A.} \bibnamefont{Minar}},
  \bibinfo{author}{\bibfnamefont{N.~C.} \bibnamefont{Cizek}},
  \bibinfo{author}{\bibfnamefont{D.}~\bibnamefont{Berryrieser}},
  \bibnamefont{and} \bibinfo{author}{\bibfnamefont{M.~A.}
  \bibnamefont{Kasevich}}, \bibinfo{journal}{Phys. Rev. Lett.}
  \textbf{\bibinfo{volume}{112}}, \bibinfo{pages}{060401}
  (\bibinfo{year}{2014}).

\bibitem[{\citenamefont{Lepoutre et~al.}(2016)\citenamefont{Lepoutre, Fouch\'e,
  Boiss\'e, Berthet, Salomon, Aspect, and Bourdel}}]{Lepoutre_sol_Ka}
\bibinfo{author}{\bibfnamefont{S.}~\bibnamefont{Lepoutre}},
  \bibinfo{author}{\bibfnamefont{L.}~\bibnamefont{Fouch\'e}},
  \bibinfo{author}{\bibfnamefont{A.}~\bibnamefont{Boiss\'e}},
  \bibinfo{author}{\bibfnamefont{G.}~\bibnamefont{Berthet}},
  \bibinfo{author}{\bibfnamefont{G.}~\bibnamefont{Salomon}},
  \bibinfo{author}{\bibfnamefont{A.}~\bibnamefont{Aspect}}, \bibnamefont{and}
  \bibinfo{author}{\bibfnamefont{T.}~\bibnamefont{Bourdel}},
  \bibinfo{journal}{Phys. Rev. A} \textbf{\bibinfo{volume}{94}},
  \bibinfo{pages}{053626} (\bibinfo{year}{2016}).

\bibitem[{\citenamefont{Boisse et~al.}({2017})\citenamefont{Boisse, Berthet,
  Fouche, Salomon, Aspect, Lepoutre, and Bourdel}}]{Boisse_disordersol_EPL}
\bibinfo{author}{\bibfnamefont{A.}~\bibnamefont{Boisse}},
  \bibinfo{author}{\bibfnamefont{G.}~\bibnamefont{Berthet}},
  \bibinfo{author}{\bibfnamefont{L.}~\bibnamefont{Fouche}},
  \bibinfo{author}{\bibfnamefont{G.}~\bibnamefont{Salomon}},
  \bibinfo{author}{\bibfnamefont{A.}~\bibnamefont{Aspect}},
  \bibinfo{author}{\bibfnamefont{S.}~\bibnamefont{Lepoutre}}, \bibnamefont{and}
  \bibinfo{author}{\bibfnamefont{T.}~\bibnamefont{Bourdel}},
  \bibinfo{journal}{{EPL}} \textbf{\bibinfo{volume}{{117}}},
  \bibinfo{pages}{{10007}} (\bibinfo{year}{{2017}}).

\bibitem[{\citenamefont{McDonald et~al.}(2014)\citenamefont{McDonald, Kuhn,
  Hardman, Bennetts, Everitt, Altin, Debs, Close, and
  Robins}}]{McDonald_solitoninterf}
\bibinfo{author}{\bibfnamefont{G.~D.} \bibnamefont{McDonald}},
  \bibinfo{author}{\bibfnamefont{C.~C.~N.} \bibnamefont{Kuhn}},
  \bibinfo{author}{\bibfnamefont{K.~S.} \bibnamefont{Hardman}},
  \bibinfo{author}{\bibfnamefont{S.}~\bibnamefont{Bennetts}},
  \bibinfo{author}{\bibfnamefont{P.~J.} \bibnamefont{Everitt}},
  \bibinfo{author}{\bibfnamefont{P.~A.} \bibnamefont{Altin}},
  \bibinfo{author}{\bibfnamefont{J.~E.} \bibnamefont{Debs}},
  \bibinfo{author}{\bibfnamefont{J.~D.} \bibnamefont{Close}}, \bibnamefont{and}
  \bibinfo{author}{\bibfnamefont{N.~P.} \bibnamefont{Robins}},
  \bibinfo{journal}{Phys. Rev. Lett.} \textbf{\bibinfo{volume}{113}},
  \bibinfo{pages}{013002} (\bibinfo{year}{2014}).

\bibitem[{\citenamefont{Everitt et~al.}(2017)\citenamefont{Everitt,
  Sooriyabandara, Guasoni, Wigley, Wei, McDonald, Hardman, Manju, Close, Kuhn
  et~al.}}]{Everitt_modinst}
\bibinfo{author}{\bibfnamefont{P.~J.} \bibnamefont{Everitt}},
  \bibinfo{author}{\bibfnamefont{M.~A.} \bibnamefont{Sooriyabandara}},
  \bibinfo{author}{\bibfnamefont{M.}~\bibnamefont{Guasoni}},
  \bibinfo{author}{\bibfnamefont{P.~B.} \bibnamefont{Wigley}},
  \bibinfo{author}{\bibfnamefont{C.~H.} \bibnamefont{Wei}},
  \bibinfo{author}{\bibfnamefont{G.~D.} \bibnamefont{McDonald}},
  \bibinfo{author}{\bibfnamefont{K.~S.} \bibnamefont{Hardman}},
  \bibinfo{author}{\bibfnamefont{P.}~\bibnamefont{Manju}},
  \bibinfo{author}{\bibfnamefont{J.~D.} \bibnamefont{Close}},
  \bibinfo{author}{\bibfnamefont{C.~C.~N.} \bibnamefont{Kuhn}},
  \bibnamefont{et~al.}, \bibinfo{journal}{Phys. Rev. A}
  \textbf{\bibinfo{volume}{96}}, \bibinfo{pages}{041601(R)}
  (\bibinfo{year}{2017}).

\bibitem[{\citenamefont{Pollack et~al.}(2009)\citenamefont{Pollack, Dries,
  Junker, Chen, Corcovilos, and Hulet}}]{Pollack_extreme_tunability_PRL}
\bibinfo{author}{\bibfnamefont{S.~E.} \bibnamefont{Pollack}},
  \bibinfo{author}{\bibfnamefont{D.}~\bibnamefont{Dries}},
  \bibinfo{author}{\bibfnamefont{M.}~\bibnamefont{Junker}},
  \bibinfo{author}{\bibfnamefont{Y.~P.} \bibnamefont{Chen}},
  \bibinfo{author}{\bibfnamefont{T.~A.} \bibnamefont{Corcovilos}},
  \bibnamefont{and} \bibinfo{author}{\bibfnamefont{R.~G.} \bibnamefont{Hulet}},
  \bibinfo{journal}{Phys. Rev. Lett.} \textbf{\bibinfo{volume}{102}},
  \bibinfo{pages}{090402} (\bibinfo{year}{2009}).

\bibitem[{\citenamefont{Me\ifmmode \check{z}\else \v{z}\fi{}nar\ifmmode
  \check{s}\else \v{s}\fi{}i\ifmmode~\check{c}\else \v{c}\fi{}
  et~al.}(2019)\citenamefont{Me\ifmmode \check{z}\else \v{z}\fi{}nar\ifmmode
  \check{s}\else \v{s}\fi{}i\ifmmode~\check{c}\else \v{c}\fi{}, Arh, Brence,
  Pi\ifmmode~\check{s}\else \v{s}\fi{}ljar, Gosar, Gosar,
  \ifmmode~\check{Z}\else \v{Z}\fi{}itko, Zupani\ifmmode~\check{c}\else
  \v{c}\fi{}, and Jegli\ifmmode~\check{c}\else
  \v{c}\fi{}}}]{Mesnarsic_cesiumsol_PhysRevA}
\bibinfo{author}{\bibfnamefont{T.}~\bibnamefont{Me\ifmmode \check{z}\else
  \v{z}\fi{}nar\ifmmode \check{s}\else \v{s}\fi{}i\ifmmode~\check{c}\else
  \v{c}\fi{}}}, \bibinfo{author}{\bibfnamefont{T.}~\bibnamefont{Arh}},
  \bibinfo{author}{\bibfnamefont{J.}~\bibnamefont{Brence}},
  \bibinfo{author}{\bibfnamefont{J.}~\bibnamefont{Pi\ifmmode~\check{s}\else
  \v{s}\fi{}ljar}}, \bibinfo{author}{\bibfnamefont{K.}~\bibnamefont{Gosar}},
  \bibinfo{author}{\bibfnamefont{i.~c.~v.} \bibnamefont{Gosar}},
  \bibinfo{author}{\bibfnamefont{R.}~\bibnamefont{\ifmmode~\check{Z}\else
  \v{Z}\fi{}itko}},
  \bibinfo{author}{\bibfnamefont{E.}~\bibnamefont{Zupani\ifmmode~\check{c}\else
  \v{c}\fi{}}}, \bibnamefont{and}
  \bibinfo{author}{\bibfnamefont{P.}~\bibnamefont{Jegli\ifmmode~\check{c}\else
  \v{c}\fi{}}}, \bibinfo{journal}{Phys. Rev. A} \textbf{\bibinfo{volume}{99}},
  \bibinfo{pages}{033625} (\bibinfo{year}{2019}).

\bibitem[{\citenamefont{Gordon}(1983)}]{gordon_forces}
\bibinfo{author}{\bibfnamefont{J.~P.} \bibnamefont{Gordon}},
  \bibinfo{journal}{Opt. Lett.} \textbf{\bibinfo{volume}{8}},
  \bibinfo{pages}{596} (\bibinfo{year}{1983}).

\bibitem[{\citenamefont{Billam and Weiss}(2014)}]{billam_NP}
\bibinfo{author}{\bibfnamefont{T.~P.} \bibnamefont{Billam}} \bibnamefont{and}
  \bibinfo{author}{\bibfnamefont{C.}~\bibnamefont{Weiss}},
  \bibinfo{journal}{Nature Physics} \textbf{\bibinfo{volume}{10}},
  \bibinfo{pages}{902} (\bibinfo{year}{2014}).

\bibitem[{\citenamefont{Al~Khawaja et~al.}(2002)\citenamefont{Al~Khawaja,
  Stoof, Hulet, Strecker, and Partridge}}]{stoof_solitons}
\bibinfo{author}{\bibfnamefont{U.}~\bibnamefont{Al~Khawaja}},
  \bibinfo{author}{\bibfnamefont{H.~T.~C.} \bibnamefont{Stoof}},
  \bibinfo{author}{\bibfnamefont{R.~G.} \bibnamefont{Hulet}},
  \bibinfo{author}{\bibfnamefont{K.~E.} \bibnamefont{Strecker}},
  \bibnamefont{and} \bibinfo{author}{\bibfnamefont{G.~B.}
  \bibnamefont{Partridge}}, \bibinfo{journal}{Phys. Rev. Lett.}
  \textbf{\bibinfo{volume}{89}}, \bibinfo{pages}{200404}
  (\bibinfo{year}{2002}).

\bibitem[{\citenamefont{D\c{a}browska-W{\"u}ster
  et~al.}(2009)\citenamefont{D\c{a}browska-W{\"u}ster, W{\"u}ster, and
  Davis}}]{wuester:collsoll}
\bibinfo{author}{\bibfnamefont{B.~J.} \bibnamefont{D\c{a}browska-W{\"u}ster}},
  \bibinfo{author}{\bibfnamefont{S.}~\bibnamefont{W{\"u}ster}},
  \bibnamefont{and} \bibinfo{author}{\bibfnamefont{M.~J.} \bibnamefont{Davis}},
  \bibinfo{journal}{New Journal of Physics} \textbf{\bibinfo{volume}{11}},
  \bibinfo{pages}{053017} (\bibinfo{year}{2009}).

\bibitem[{\citenamefont{Carr and Brand}(2004)}]{brand_solitons}
\bibinfo{author}{\bibfnamefont{L.~D.} \bibnamefont{Carr}} \bibnamefont{and}
  \bibinfo{author}{\bibfnamefont{J.}~\bibnamefont{Brand}},
  \bibinfo{journal}{Phys. Rev. Lett.} \textbf{\bibinfo{volume}{92}},
  \bibinfo{pages}{040401} (\bibinfo{year}{2004}).

\bibitem[{\citenamefont{Salasnich et~al.}(2003)\citenamefont{Salasnich, Parola,
  and Reatto}}]{Salasnich_modinst}
\bibinfo{author}{\bibfnamefont{L.}~\bibnamefont{Salasnich}},
  \bibinfo{author}{\bibfnamefont{A.}~\bibnamefont{Parola}}, \bibnamefont{and}
  \bibinfo{author}{\bibfnamefont{L.}~\bibnamefont{Reatto}},
  \bibinfo{journal}{Phys. Rev. Lett.} \textbf{\bibinfo{volume}{91}},
  \bibinfo{pages}{080405} (\bibinfo{year}{2003}).

\bibitem[{\citenamefont{Streltsov
  et~al.}(2011{\natexlab{a}})\citenamefont{Streltsov, Alon, and
  Cederbaum}}]{streltsov_frag}
\bibinfo{author}{\bibfnamefont{A.~I.} \bibnamefont{Streltsov}},
  \bibinfo{author}{\bibfnamefont{O.~E.} \bibnamefont{Alon}}, \bibnamefont{and}
  \bibinfo{author}{\bibfnamefont{L.~S.} \bibnamefont{Cederbaum}},
  \bibinfo{journal}{Phys. Rev. Lett.} \textbf{\bibinfo{volume}{106}},
  \bibinfo{pages}{240401} (\bibinfo{year}{2011}{\natexlab{a}}).

\bibitem[{\citenamefont{Lewenstein and You}(1996)}]{lewenstein_phasediff}
\bibinfo{author}{\bibfnamefont{M.}~\bibnamefont{Lewenstein}} \bibnamefont{and}
  \bibinfo{author}{\bibfnamefont{L.}~\bibnamefont{You}},
  \bibinfo{journal}{Phys. Rev. Lett.} \textbf{\bibinfo{volume}{77}},
  \bibinfo{pages}{3489} (\bibinfo{year}{1996}).

\bibitem[{\citenamefont{Albiez et~al.}(2005)\citenamefont{Albiez, Gati,
  F\"olling, Hunsmann, Cristiani, and Oberthaler}}]{Albiez:oberthaler:BJJ}
\bibinfo{author}{\bibfnamefont{M.}~\bibnamefont{Albiez}},
  \bibinfo{author}{\bibfnamefont{R.}~\bibnamefont{Gati}},
  \bibinfo{author}{\bibfnamefont{J.}~\bibnamefont{F\"olling}},
  \bibinfo{author}{\bibfnamefont{S.}~\bibnamefont{Hunsmann}},
  \bibinfo{author}{\bibfnamefont{M.}~\bibnamefont{Cristiani}},
  \bibnamefont{and} \bibinfo{author}{\bibfnamefont{M.~K.}
  \bibnamefont{Oberthaler}}, \bibinfo{journal}{Phys. Rev. Lett.}
  \textbf{\bibinfo{volume}{95}}, \bibinfo{pages}{010402}
  (\bibinfo{year}{2005}).

\bibitem[{\citenamefont{Parker et~al.}(2008)\citenamefont{Parker, Martin,
  Cornish, and Adams}}]{parker_bsw}
\bibinfo{author}{\bibfnamefont{N.~G.} \bibnamefont{Parker}},
  \bibinfo{author}{\bibfnamefont{A.~M.} \bibnamefont{Martin}},
  \bibinfo{author}{\bibfnamefont{S.~L.} \bibnamefont{Cornish}},
  \bibnamefont{and} \bibinfo{author}{\bibfnamefont{C.~S.} \bibnamefont{Adams}},
  \bibinfo{journal}{J. Phys. B: At. Mol. Opt. Phys.}
  \textbf{\bibinfo{volume}{41}}, \bibinfo{pages}{045303}
  (\bibinfo{year}{2008}).

\bibitem[{\citenamefont{Mitschke and Mollenhauer}(1987)}]{mitschke_forces}
\bibinfo{author}{\bibfnamefont{F.~M.} \bibnamefont{Mitschke}} \bibnamefont{and}
  \bibinfo{author}{\bibfnamefont{L.~F.} \bibnamefont{Mollenhauer}},
  \bibinfo{journal}{Opt. Lett.} \textbf{\bibinfo{volume}{12}},
  \bibinfo{pages}{355} (\bibinfo{year}{1987}).

\bibitem[{\citenamefont{Dodd et~al.}(1996)\citenamefont{Dodd, Edwards,
  Williams, Clark, Holland, Ruprecht, and Burnett}}]{ruprecht:attractive}
\bibinfo{author}{\bibfnamefont{R.~J.} \bibnamefont{Dodd}},
  \bibinfo{author}{\bibfnamefont{M.}~\bibnamefont{Edwards}},
  \bibinfo{author}{\bibfnamefont{C.~J.} \bibnamefont{Williams}},
  \bibinfo{author}{\bibfnamefont{C.~W.} \bibnamefont{Clark}},
  \bibinfo{author}{\bibfnamefont{M.~J.} \bibnamefont{Holland}},
  \bibinfo{author}{\bibfnamefont{P.~A.} \bibnamefont{Ruprecht}},
  \bibnamefont{and} \bibinfo{author}{\bibfnamefont{K.}~\bibnamefont{Burnett}},
  \bibinfo{journal}{Phys. Rev. A} \textbf{\bibinfo{volume}{54}},
  \bibinfo{pages}{661} (\bibinfo{year}{1996}).

\bibitem[{par(2009)}]{parker_previouspreprint}
\bibinfo{journal}{Physica D: Nonlinear Phenomena}
  \textbf{\bibinfo{volume}{238}}, \bibinfo{pages}{1456} (\bibinfo{year}{2009}).

\bibitem[{\citenamefont{Parker et~al.}(2007)\citenamefont{Parker, Cornish,
  Adams, and Martin}}]{parker_bsw1}
\bibinfo{author}{\bibfnamefont{N.~G.} \bibnamefont{Parker}},
  \bibinfo{author}{\bibfnamefont{S.~L.} \bibnamefont{Cornish}},
  \bibinfo{author}{\bibfnamefont{C.~S.} \bibnamefont{Adams}}, \bibnamefont{and}
  \bibinfo{author}{\bibfnamefont{A.~M.} \bibnamefont{Martin}},
  \bibinfo{journal}{J. Phys. B: At. Mol. Opt. Phys.}
  \textbf{\bibinfo{volume}{40}}, \bibinfo{pages}{3127} (\bibinfo{year}{2007}).

\bibitem[{\citenamefont{Leung et~al.}(2002)\citenamefont{Leung, Truscott, and
  Baldwin}}]{li_theory}
\bibinfo{author}{\bibfnamefont{V.~Y.~F.} \bibnamefont{Leung}},
  \bibinfo{author}{\bibfnamefont{A.~G.} \bibnamefont{Truscott}},
  \bibnamefont{and} \bibinfo{author}{\bibfnamefont{K.~G.~H.}
  \bibnamefont{Baldwin}}, \bibinfo{journal}{Phys. Rev. A}
  \textbf{\bibinfo{volume}{66}}, \bibinfo{pages}{061602(R)}
  (\bibinfo{year}{2002}).

\bibitem[{\citenamefont{Alon et~al.}(2008)\citenamefont{Alon, Streltsov, and
  Cederbaum}}]{alon:pra:mctdhb}
\bibinfo{author}{\bibfnamefont{O.~E.} \bibnamefont{Alon}},
  \bibinfo{author}{\bibfnamefont{A.~I.} \bibnamefont{Streltsov}},
  \bibnamefont{and} \bibinfo{author}{\bibfnamefont{L.~S.}
  \bibnamefont{Cederbaum}}, \bibinfo{journal}{Phys. Rev. A}
  \textbf{\bibinfo{volume}{77}}, \bibinfo{pages}{033613}
  (\bibinfo{year}{2008}).

\bibitem[{\citenamefont{Streltsov et~al.}(2008)\citenamefont{Streltsov, Alon,
  and Cederbaum}}]{Streltsov_fragmentation_PRL2008}
\bibinfo{author}{\bibfnamefont{A.~I.} \bibnamefont{Streltsov}},
  \bibinfo{author}{\bibfnamefont{O.~E.} \bibnamefont{Alon}}, \bibnamefont{and}
  \bibinfo{author}{\bibfnamefont{L.~S.} \bibnamefont{Cederbaum}},
  \bibinfo{journal}{Phys. Rev. Lett.} \textbf{\bibinfo{volume}{100}},
  \bibinfo{pages}{130401} (\bibinfo{year}{2008}).

\bibitem[{\citenamefont{Alon and Cederbaum}(2018)}]{alon:attractive:anharmonic}
\bibinfo{author}{\bibfnamefont{O.~E.} \bibnamefont{Alon}} \bibnamefont{and}
  \bibinfo{author}{\bibfnamefont{L.~S.} \bibnamefont{Cederbaum}},
  \bibinfo{journal}{Chem. Phys.} \textbf{\bibinfo{volume}{515}},
  \bibinfo{pages}{287} (\bibinfo{year}{2018}).

\bibitem[{\citenamefont{Sakmann et~al.}(2014)\citenamefont{Sakmann, Streltsov,
  Alon, and Cederbaum}}]{Sakmann_universalfrag_PhysRevA}
\bibinfo{author}{\bibfnamefont{K.}~\bibnamefont{Sakmann}},
  \bibinfo{author}{\bibfnamefont{A.~I.} \bibnamefont{Streltsov}},
  \bibinfo{author}{\bibfnamefont{O.~E.} \bibnamefont{Alon}}, \bibnamefont{and}
  \bibinfo{author}{\bibfnamefont{L.~S.} \bibnamefont{Cederbaum}},
  \bibinfo{journal}{Phys. Rev. A} \textbf{\bibinfo{volume}{89}},
  \bibinfo{pages}{023602} (\bibinfo{year}{2014}).

\bibitem[{\citenamefont{Katsimiga et~al.}(2018)\citenamefont{Katsimiga,
  Mistakidis, Koutentakis, Kevrekidis, and
  Schmelcher}}]{Katsimiga_impurity_PRA}
\bibinfo{author}{\bibfnamefont{G.~C.} \bibnamefont{Katsimiga}},
  \bibinfo{author}{\bibfnamefont{S.~I.} \bibnamefont{Mistakidis}},
  \bibinfo{author}{\bibfnamefont{G.~M.} \bibnamefont{Koutentakis}},
  \bibinfo{author}{\bibfnamefont{P.~G.} \bibnamefont{Kevrekidis}},
  \bibnamefont{and}
  \bibinfo{author}{\bibfnamefont{P.}~\bibnamefont{Schmelcher}},
  \bibinfo{journal}{Phys. Rev. A} \textbf{\bibinfo{volume}{98}},
  \bibinfo{pages}{013632} (\bibinfo{year}{2018}).

\bibitem[{\citenamefont{Katsimiga
  et~al.}(2017{\natexlab{a}})\citenamefont{Katsimiga, Mistakidis, Koutentakis,
  Kevrekidis, and Schmelcher}}]{Katsimiga_bentdark_NJP2017}
\bibinfo{author}{\bibfnamefont{G.~C.} \bibnamefont{Katsimiga}},
  \bibinfo{author}{\bibfnamefont{S.~I.} \bibnamefont{Mistakidis}},
  \bibinfo{author}{\bibfnamefont{G.~M.} \bibnamefont{Koutentakis}},
  \bibinfo{author}{\bibfnamefont{P.~G.} \bibnamefont{Kevrekidis}},
  \bibnamefont{and}
  \bibinfo{author}{\bibfnamefont{P.}~\bibnamefont{Schmelcher}},
  \bibinfo{journal}{New J. Phys.} \textbf{\bibinfo{volume}{19}},
  \bibinfo{pages}{123012} (\bibinfo{year}{2017}{\natexlab{a}}).

\bibitem[{\citenamefont{Streltsov et~al.}(2009)\citenamefont{Streltsov, Alon,
  and Cederbaum}}]{Streltsov_barrier_PRA2009}
\bibinfo{author}{\bibfnamefont{A.~I.} \bibnamefont{Streltsov}},
  \bibinfo{author}{\bibfnamefont{O.~E.} \bibnamefont{Alon}}, \bibnamefont{and}
  \bibinfo{author}{\bibfnamefont{L.~S.} \bibnamefont{Cederbaum}},
  \bibinfo{journal}{Phys. Rev. A} \textbf{\bibinfo{volume}{80}},
  \bibinfo{pages}{043616} (\bibinfo{year}{2009}).

\bibitem[{\citenamefont{Grond et~al.}(2011)\citenamefont{Grond, Betz,
  Hohenester, Mauser, Schmiedmayer, and Schumm}}]{Grond_Shapiro_NJP2011}
\bibinfo{author}{\bibfnamefont{J.}~\bibnamefont{Grond}},
  \bibinfo{author}{\bibfnamefont{T.}~\bibnamefont{Betz}},
  \bibinfo{author}{\bibfnamefont{U.}~\bibnamefont{Hohenester}},
  \bibinfo{author}{\bibfnamefont{N.~J.} \bibnamefont{Mauser}},
  \bibinfo{author}{\bibfnamefont{J.}~\bibnamefont{Schmiedmayer}},
  \bibnamefont{and} \bibinfo{author}{\bibfnamefont{T.}~\bibnamefont{Schumm}},
  \bibinfo{journal}{New J. Phys.} \textbf{\bibinfo{volume}{13}},
  \bibinfo{pages}{065026} (\bibinfo{year}{2011}).

\bibitem[{\citenamefont{Kr{\"o}nke et~al.}(2015)\citenamefont{Kr{\"o}nke,
  Kn{\"o}rzer, and Schmelcher}}]{Kroenke_impurity_NJP2015}
\bibinfo{author}{\bibfnamefont{S.}~\bibnamefont{Kr{\"o}nke}},
  \bibinfo{author}{\bibfnamefont{J.}~\bibnamefont{Kn{\"o}rzer}},
  \bibnamefont{and}
  \bibinfo{author}{\bibfnamefont{P.}~\bibnamefont{Schmelcher}},
  \bibinfo{journal}{New J. Phys.} \textbf{\bibinfo{volume}{17}},
  \bibinfo{pages}{053001} (\bibinfo{year}{2015}).

\bibitem[{\citenamefont{Katsimiga
  et~al.}(2017{\natexlab{b}})\citenamefont{Katsimiga, Koutentakis, Mistakidis,
  Kevrekidis, and Schmelcher}}]{Katsimiga_darkbright_NJP2017}
\bibinfo{author}{\bibfnamefont{G.~C.} \bibnamefont{Katsimiga}},
  \bibinfo{author}{\bibfnamefont{G.~M.} \bibnamefont{Koutentakis}},
  \bibinfo{author}{\bibfnamefont{S.~I.} \bibnamefont{Mistakidis}},
  \bibinfo{author}{\bibfnamefont{P.~G.} \bibnamefont{Kevrekidis}},
  \bibnamefont{and}
  \bibinfo{author}{\bibfnamefont{P.}~\bibnamefont{Schmelcher}},
  \bibinfo{journal}{New J. Phys.} \textbf{\bibinfo{volume}{19}},
  \bibinfo{pages}{073004} (\bibinfo{year}{2017}{\natexlab{b}}).

\bibitem[{\citenamefont{Cosme et~al.}(2017)\citenamefont{Cosme, Andersen, and
  Brand}}]{cosme_intblock_pra}
\bibinfo{author}{\bibfnamefont{J.~G.} \bibnamefont{Cosme}},
  \bibinfo{author}{\bibfnamefont{M.~F.} \bibnamefont{Andersen}},
  \bibnamefont{and} \bibinfo{author}{\bibfnamefont{J.}~\bibnamefont{Brand}},
  \bibinfo{journal}{Phys. Rev. A} \textbf{\bibinfo{volume}{96}},
  \bibinfo{pages}{013616} (\bibinfo{year}{2017}).

\bibitem[{\citenamefont{Sakmann et~al.}(2012)\citenamefont{Sakmann, Lode,
  Streltsov, Alon, and Cederbaum}}]{open_mctdhb}
\bibinfo{author}{\bibfnamefont{K.}~\bibnamefont{Sakmann}},
  \bibinfo{author}{\bibfnamefont{A.~U.~J.} \bibnamefont{Lode}},
  \bibinfo{author}{\bibfnamefont{A.~I.} \bibnamefont{Streltsov}},
  \bibinfo{author}{\bibfnamefont{O.~E.} \bibnamefont{Alon}}, \bibnamefont{and}
  \bibinfo{author}{\bibfnamefont{L.~S.} \bibnamefont{Cederbaum}},
  \emph{\bibinfo{title}{Openmctdhb v2.3}} (\bibinfo{year}{2012}),
  \bibinfo{note}{http://OpenMCTDHB.uni-hd.de}.

\bibitem[{\citenamefont{Steel et~al.}(1998)\citenamefont{Steel, Olsen, Plimak,
  Drummond, Tan, Collett, Walls, and Graham}}]{steel:wigner}
\bibinfo{author}{\bibfnamefont{M.~J.} \bibnamefont{Steel}},
  \bibinfo{author}{\bibfnamefont{M.~K.} \bibnamefont{Olsen}},
  \bibinfo{author}{\bibfnamefont{L.~I.} \bibnamefont{Plimak}},
  \bibinfo{author}{\bibfnamefont{P.~D.} \bibnamefont{Drummond}},
  \bibinfo{author}{\bibfnamefont{S.~M.} \bibnamefont{Tan}},
  \bibinfo{author}{\bibfnamefont{M.~J.} \bibnamefont{Collett}},
  \bibinfo{author}{\bibfnamefont{D.~F.} \bibnamefont{Walls}}, \bibnamefont{and}
  \bibinfo{author}{\bibfnamefont{R.}~\bibnamefont{Graham}},
  \bibinfo{journal}{Phys. Rev. A} \textbf{\bibinfo{volume}{58}},
  \bibinfo{pages}{4824} (\bibinfo{year}{1998}).

\bibitem[{\citenamefont{Sinatra et~al.}(2001)\citenamefont{Sinatra, Lobo, and
  Castin}}]{Sinatra2001}
\bibinfo{author}{\bibfnamefont{A.}~\bibnamefont{Sinatra}},
  \bibinfo{author}{\bibfnamefont{C.}~\bibnamefont{Lobo}}, \bibnamefont{and}
  \bibinfo{author}{\bibfnamefont{Y.}~\bibnamefont{Castin}},
  \bibinfo{journal}{Phys. Rev. Lett.} \textbf{\bibinfo{volume}{87}},
  \bibinfo{pages}{210404} (\bibinfo{year}{2001}).

\bibitem[{\citenamefont{Sinatra et~al.}(2002)\citenamefont{Sinatra, Lobo, and
  Castin}}]{castin:validity}
\bibinfo{author}{\bibfnamefont{A.}~\bibnamefont{Sinatra}},
  \bibinfo{author}{\bibfnamefont{C.}~\bibnamefont{Lobo}}, \bibnamefont{and}
  \bibinfo{author}{\bibfnamefont{Y.}~\bibnamefont{Castin}},
  \bibinfo{journal}{J. Phys. B: At. Mol. Opt. Phys.}
  \textbf{\bibinfo{volume}{35}}, \bibinfo{pages}{3599} (\bibinfo{year}{2002}).

\bibitem[{\citenamefont{Blakie et~al.}(2008)\citenamefont{Blakie, Bradley,
  Davis, Ballagh, and Gardiner}}]{blair:review}
\bibinfo{author}{\bibfnamefont{P.}~\bibnamefont{Blakie}},
  \bibinfo{author}{\bibfnamefont{A.}~\bibnamefont{Bradley}},
  \bibinfo{author}{\bibfnamefont{M.}~\bibnamefont{Davis}},
  \bibinfo{author}{\bibfnamefont{R.}~\bibnamefont{Ballagh}}, \bibnamefont{and}
  \bibinfo{author}{\bibfnamefont{C.}~\bibnamefont{Gardiner}},
  \bibinfo{journal}{Advances in Physics} \textbf{\bibinfo{volume}{57}},
  \bibinfo{pages}{363} (\bibinfo{year}{2008}).

\bibitem[{\citenamefont{Hush et~al.}(2010)\citenamefont{Hush, Carvalho, and
  Hope}}]{hush_numberphasewigner}
\bibinfo{author}{\bibfnamefont{M.~R.} \bibnamefont{Hush}},
  \bibinfo{author}{\bibfnamefont{A.~R.~R.} \bibnamefont{Carvalho}},
  \bibnamefont{and} \bibinfo{author}{\bibfnamefont{J.~J.} \bibnamefont{Hope}},
  \bibinfo{journal}{Phys. Rev. A} \textbf{\bibinfo{volume}{81}},
  \bibinfo{pages}{033852} (\bibinfo{year}{2010}).

\bibitem[{\citenamefont{Corney and Olsen}(2015)}]{corney_nongauss_posp_wig}
\bibinfo{author}{\bibfnamefont{J.~F.} \bibnamefont{Corney}} \bibnamefont{and}
  \bibinfo{author}{\bibfnamefont{M.~K.} \bibnamefont{Olsen}},
  \bibinfo{journal}{Phys. Rev. A} \textbf{\bibinfo{volume}{91}},
  \bibinfo{pages}{023824} (\bibinfo{year}{2015}).

\bibitem[{\citenamefont{Polkovnikov}(2003)}]{polkovnikov:timescale}
\bibinfo{author}{\bibfnamefont{A.}~\bibnamefont{Polkovnikov}},
  \bibinfo{journal}{Phys. Rev. A} \textbf{\bibinfo{volume}{68}},
  \bibinfo{pages}{033609} (\bibinfo{year}{2003}).

\bibitem[{\citenamefont{Norrie et~al.}(2005)\citenamefont{Norrie, Ballagh, and
  Gardiner}}]{norrie:prl}
\bibinfo{author}{\bibfnamefont{A.~A.} \bibnamefont{Norrie}},
  \bibinfo{author}{\bibfnamefont{R.~J.} \bibnamefont{Ballagh}},
  \bibnamefont{and} \bibinfo{author}{\bibfnamefont{C.~W.}
  \bibnamefont{Gardiner}}, \bibinfo{journal}{Phys. Rev. Lett.}
  \textbf{\bibinfo{volume}{94}}, \bibinfo{pages}{040401}
  (\bibinfo{year}{2005}).

\bibitem[{\citenamefont{Norrie et~al.}(2006)\citenamefont{Norrie, Ballagh, and
  Gardiner}}]{norrie:long}
\bibinfo{author}{\bibfnamefont{A.~A.} \bibnamefont{Norrie}},
  \bibinfo{author}{\bibfnamefont{R.~J.} \bibnamefont{Ballagh}},
  \bibnamefont{and} \bibinfo{author}{\bibfnamefont{C.~W.}
  \bibnamefont{Gardiner}}, \bibinfo{journal}{Phys. Rev. A}
  \textbf{\bibinfo{volume}{73}}, \bibinfo{pages}{043617}
  (\bibinfo{year}{2006}).

\bibitem[{\citenamefont{Norrie}(2005)}]{norrie:thesis}
\bibinfo{author}{\bibfnamefont{A.~A.} \bibnamefont{Norrie}}, Ph.D. thesis,
  \bibinfo{school}{University of Otago} (\bibinfo{year}{2005}).

\bibitem[{\citenamefont{Dennis et~al.}(2012)\citenamefont{Dennis, Hope, and
  Johnsson}}]{xmds:docu}
\bibinfo{author}{\bibfnamefont{G.~R.} \bibnamefont{Dennis}},
  \bibinfo{author}{\bibfnamefont{J.~J.} \bibnamefont{Hope}}, \bibnamefont{and}
  \bibinfo{author}{\bibfnamefont{M.~T.} \bibnamefont{Johnsson}}
  (\bibinfo{year}{2012}), \bibinfo{note}{http://www.xmds.org/}.

\bibitem[{\citenamefont{Dennis et~al.}(2013)\citenamefont{Dennis, Hope, and
  Johnsson}}]{xmds:paper}
\bibinfo{author}{\bibfnamefont{G.~R.} \bibnamefont{Dennis}},
  \bibinfo{author}{\bibfnamefont{J.~J.} \bibnamefont{Hope}}, \bibnamefont{and}
  \bibinfo{author}{\bibfnamefont{M.~T.} \bibnamefont{Johnsson}},
  \bibinfo{journal}{Comput. Phys. Comm.} \textbf{\bibinfo{volume}{184}},
  \bibinfo{pages}{201} (\bibinfo{year}{2013}).

\bibitem[{\citenamefont{Penrose and Onsager}(1956)}]{penrose_onsager_crit}
\bibinfo{author}{\bibfnamefont{O.}~\bibnamefont{Penrose}} \bibnamefont{and}
  \bibinfo{author}{\bibfnamefont{L.}~\bibnamefont{Onsager}},
  \bibinfo{journal}{Phys. Rev. A} \textbf{\bibinfo{volume}{104}},
  \bibinfo{pages}{576} (\bibinfo{year}{1956}).

\bibitem[{\citenamefont{Blakie and Davis}(2005)}]{Blakie2005}
\bibinfo{author}{\bibfnamefont{P.~B.} \bibnamefont{Blakie}} \bibnamefont{and}
  \bibinfo{author}{\bibfnamefont{M.~J.} \bibnamefont{Davis}},
  \bibinfo{journal}{Phys. Rev. A} \textbf{\bibinfo{volume}{72}},
  \bibinfo{pages}{063608} (\bibinfo{year}{2005}).

\bibitem[{\citenamefont{Streltsov
  et~al.}(2011{\natexlab{b}})\citenamefont{Streltsov, Sakmann, Alon, and
  Cederbaum}}]{streltsov:triplewellMCTDHB:PhysRevA.83.043604}
\bibinfo{author}{\bibfnamefont{A.~I.} \bibnamefont{Streltsov}},
  \bibinfo{author}{\bibfnamefont{K.}~\bibnamefont{Sakmann}},
  \bibinfo{author}{\bibfnamefont{O.~E.} \bibnamefont{Alon}}, \bibnamefont{and}
  \bibinfo{author}{\bibfnamefont{L.~S.} \bibnamefont{Cederbaum}},
  \bibinfo{journal}{Phys. Rev. A} \textbf{\bibinfo{volume}{83}},
  \bibinfo{pages}{043604} (\bibinfo{year}{2011}{\natexlab{b}}).

\bibitem[{\citenamefont{Walls and Milburn}(1994)}]{book:walls:milburn}
\bibinfo{author}{\bibfnamefont{D.~F.} \bibnamefont{Walls}} \bibnamefont{and}
  \bibinfo{author}{\bibfnamefont{G.~J.} \bibnamefont{Milburn}},
  \emph{\bibinfo{title}{Quantum Optics}} (\bibinfo{publisher}{Springer Verlag},
  \bibinfo{year}{1994}).

\bibitem[{\citenamefont{Johnsson and Haine}(2007)}]{matthias:simon:kerr}
\bibinfo{author}{\bibfnamefont{M.~T.} \bibnamefont{Johnsson}} \bibnamefont{and}
  \bibinfo{author}{\bibfnamefont{S.~A.} \bibnamefont{Haine}},
  \bibinfo{journal}{Phys. Rev. Lett.} \textbf{\bibinfo{volume}{99}},
  \bibinfo{pages}{010401} (\bibinfo{year}{2007}).

\bibitem[{\citenamefont{W{\"u}ster et~al.}(2008)\citenamefont{W{\"u}ster,
  D\c{a}browska-W{\"u}ster, Scott, Close, and Savage}}]{wuester:kerr}
\bibinfo{author}{\bibfnamefont{S.}~\bibnamefont{W{\"u}ster}},
  \bibinfo{author}{\bibfnamefont{B.~J.}
  \bibnamefont{D\c{a}browska-W{\"u}ster}},
  \bibinfo{author}{\bibfnamefont{S.~M.} \bibnamefont{Scott}},
  \bibinfo{author}{\bibfnamefont{J.~D.} \bibnamefont{Close}}, \bibnamefont{and}
  \bibinfo{author}{\bibfnamefont{C.~M.} \bibnamefont{Savage}},
  \bibinfo{journal}{Phys. Rev. A} \textbf{\bibinfo{volume}{77}},
  \bibinfo{pages}{023619} (\bibinfo{year}{2008}).

\bibitem[{\citenamefont{Lai and Haus}(1989{\natexlab{a}})}]{lai_quantsol_I}
\bibinfo{author}{\bibfnamefont{Y.}~\bibnamefont{Lai}} \bibnamefont{and}
  \bibinfo{author}{\bibfnamefont{H.~A.} \bibnamefont{Haus}},
  \bibinfo{journal}{Phys. Rev. A} \textbf{\bibinfo{volume}{40}},
  \bibinfo{pages}{844} (\bibinfo{year}{1989}{\natexlab{a}}).

\bibitem[{\citenamefont{Lai and Haus}(1989{\natexlab{b}})}]{lai_quantsol_II}
\bibinfo{author}{\bibfnamefont{Y.}~\bibnamefont{Lai}} \bibnamefont{and}
  \bibinfo{author}{\bibfnamefont{H.~A.} \bibnamefont{Haus}},
  \bibinfo{journal}{Phys. Rev. A} \textbf{\bibinfo{volume}{40}},
  \bibinfo{pages}{854} (\bibinfo{year}{1989}{\natexlab{b}}).

\bibitem[{\citenamefont{Martin and
  Ruostekoski}(2012)}]{Martin_solinterf_NJP_2012}
\bibinfo{author}{\bibfnamefont{A.~D.} \bibnamefont{Martin}} \bibnamefont{and}
  \bibinfo{author}{\bibfnamefont{J.}~\bibnamefont{Ruostekoski}},
  \bibinfo{journal}{New J. Phys.} \textbf{\bibinfo{volume}{14}},
  \bibinfo{pages}{043040} (\bibinfo{year}{2012}).

\bibitem[{\citenamefont{Sakmann and Kasevich}(2016)}]{sakman_singleshot_nphys}
\bibinfo{author}{\bibfnamefont{K.}~\bibnamefont{Sakmann}} \bibnamefont{and}
  \bibinfo{author}{\bibfnamefont{M.}~\bibnamefont{Kasevich}},
  \bibinfo{journal}{Nature Physics} \textbf{\bibinfo{volume}{12}},
  \bibinfo{pages}{451} (\bibinfo{year}{2016}).

\bibitem[{\citenamefont{Kevrekidis et~al.}(2005)\citenamefont{Kevrekidis,
  Frantzeskakis, Carretero-Gonz\'alez, Malomed, Herring, and
  Bishop}}]{Kevrekidis_radiation_PhysRevA}
\bibinfo{author}{\bibfnamefont{P.~G.} \bibnamefont{Kevrekidis}},
  \bibinfo{author}{\bibfnamefont{D.~J.} \bibnamefont{Frantzeskakis}},
  \bibinfo{author}{\bibfnamefont{R.}~\bibnamefont{Carretero-Gonz\'alez}},
  \bibinfo{author}{\bibfnamefont{B.~A.} \bibnamefont{Malomed}},
  \bibinfo{author}{\bibfnamefont{G.}~\bibnamefont{Herring}}, \bibnamefont{and}
  \bibinfo{author}{\bibfnamefont{A.~R.} \bibnamefont{Bishop}},
  \bibinfo{journal}{Phys. Rev. A} \textbf{\bibinfo{volume}{71}},
  \bibinfo{pages}{023614} (\bibinfo{year}{2005}).

\bibitem[{\citenamefont{Lewenstein and
  Malomed}(2009)}]{Lewenstein_phasekin_entangle}
\bibinfo{author}{\bibfnamefont{M.}~\bibnamefont{Lewenstein}} \bibnamefont{and}
  \bibinfo{author}{\bibfnamefont{B.~A.} \bibnamefont{Malomed}},
  \bibinfo{journal}{New Journal of Physics} \textbf{\bibinfo{volume}{11}},
  \bibinfo{pages}{113014} (\bibinfo{year}{2009}).

\bibitem[{\citenamefont{Gertjerenken et~al.}(2013)\citenamefont{Gertjerenken,
  Billam, Blackley, Le~Sueur, Khaykovich, Cornish, and
  Weiss}}]{Gertjerenken_cat_coll_PRL}
\bibinfo{author}{\bibfnamefont{B.}~\bibnamefont{Gertjerenken}},
  \bibinfo{author}{\bibfnamefont{T.~P.} \bibnamefont{Billam}},
  \bibinfo{author}{\bibfnamefont{C.~L.} \bibnamefont{Blackley}},
  \bibinfo{author}{\bibfnamefont{C.~R.} \bibnamefont{Le~Sueur}},
  \bibinfo{author}{\bibfnamefont{L.}~\bibnamefont{Khaykovich}},
  \bibinfo{author}{\bibfnamefont{S.~L.} \bibnamefont{Cornish}},
  \bibnamefont{and} \bibinfo{author}{\bibfnamefont{C.}~\bibnamefont{Weiss}},
  \bibinfo{journal}{Phys. Rev. Lett.} \textbf{\bibinfo{volume}{111}},
  \bibinfo{pages}{100406} (\bibinfo{year}{2013}).

\bibitem[{\citenamefont{Mishmash and
  Carr}(2009)}]{Mishmash_entangleddarksol_PRL}
\bibinfo{author}{\bibfnamefont{R.~V.} \bibnamefont{Mishmash}} \bibnamefont{and}
  \bibinfo{author}{\bibfnamefont{L.~D.} \bibnamefont{Carr}},
  \bibinfo{journal}{Phys. Rev. Lett.} \textbf{\bibinfo{volume}{103}},
  \bibinfo{pages}{140403} (\bibinfo{year}{2009}).

\bibitem[{\citenamefont{McGuire}(1964)}]{McGuire_exactlysolvable_JMP}
\bibinfo{author}{\bibfnamefont{J.~B.} \bibnamefont{McGuire}},
  \bibinfo{journal}{Journal of Mathematical Physics}
  \textbf{\bibinfo{volume}{5}}, \bibinfo{pages}{622} (\bibinfo{year}{1964}).

\bibitem[{\citenamefont{Lieb and Liniger}(1963)}]{LL_model_PR}
\bibinfo{author}{\bibfnamefont{E.~H.} \bibnamefont{Lieb}} \bibnamefont{and}
  \bibinfo{author}{\bibfnamefont{W.}~\bibnamefont{Liniger}},
  \bibinfo{journal}{Phys. Rev.} \textbf{\bibinfo{volume}{130}},
  \bibinfo{pages}{1605} (\bibinfo{year}{1963}).

\bibitem[{\citenamefont{Holdaway et~al.}(2014)\citenamefont{Holdaway, Weiss,
  and Gardiner}}]{Holdaway_entanglesol}
\bibinfo{author}{\bibfnamefont{D.~I.~H.} \bibnamefont{Holdaway}},
  \bibinfo{author}{\bibfnamefont{C.}~\bibnamefont{Weiss}}, \bibnamefont{and}
  \bibinfo{author}{\bibfnamefont{S.~A.} \bibnamefont{Gardiner}},
  \bibinfo{journal}{Phys. Rev. A} \textbf{\bibinfo{volume}{89}},
  \bibinfo{pages}{013611} (\bibinfo{year}{2014}).

\bibitem[{\citenamefont{{Yu-Zhu} et~al.}(2015)\citenamefont{{Yu-Zhu},
  {Yang-Yang}, and {Xi-Wen}}}]{zhu_manybody_liebliniger_ChinPhysB}
\bibinfo{author}{\bibfnamefont{J.}~\bibnamefont{{Yu-Zhu}}},
  \bibinfo{author}{\bibfnamefont{C.}~\bibnamefont{{Yang-Yang}}},
  \bibnamefont{and} \bibinfo{author}{\bibfnamefont{G.}~\bibnamefont{{Xi-Wen}}},
  \bibinfo{journal}{Chinese Physics B} \textbf{\bibinfo{volume}{24}},
  \bibinfo{pages}{050311} (\bibinfo{year}{2015}).
	
\bibitem[{\citenamefont{Kinoshita et~al.}(2006)\citenamefont{Kinoshita, Wenger,
  and Weiss}}]{kinoshita_cradle}
\bibinfo{author}{\bibfnamefont{T.}~\bibnamefont{Kinoshita}},
  \bibinfo{author}{\bibfnamefont{T.}~\bibnamefont{Wenger}}, \bibnamefont{and}
  \bibinfo{author}{\bibfnamefont{D.~S.} \bibnamefont{Weiss}},
  \bibinfo{journal}{Nature} \textbf{\bibinfo{volume}{440}},
  \bibinfo{pages}{900} (\bibinfo{year}{2006}).

\bibitem[{\citenamefont{Hofferberth et~al.}(2008)\citenamefont{Hofferberth,
  Lesanovsky, Fischer, Schumm, Imambekov, Gritsev, Demler, and
  Schmiedmayer}}]{hofferberth_quantum_thermal}
\bibinfo{author}{\bibfnamefont{S.}~\bibnamefont{Hofferberth}},
  \bibinfo{author}{\bibfnamefont{I.}~\bibnamefont{Lesanovsky}},
  \bibinfo{author}{\bibfnamefont{B.}~\bibnamefont{Fischer}},
  \bibinfo{author}{\bibfnamefont{T.}~\bibnamefont{Schumm}},
  \bibinfo{author}{\bibfnamefont{A.}~\bibnamefont{Imambekov}},
  \bibinfo{author}{\bibfnamefont{V.}~\bibnamefont{Gritsev}},
  \bibinfo{author}{\bibfnamefont{E.}~\bibnamefont{Demler}}, \bibnamefont{and}
  \bibinfo{author}{\bibfnamefont{J.}~\bibnamefont{Schmiedmayer}},
  \bibinfo{journal}{Nature Physics} \textbf{\bibinfo{volume}{4}},
  \bibinfo{pages}{489} (\bibinfo{year}{2008}).

\bibitem[{\citenamefont{Hofferberth et~al.}(2007)\citenamefont{Hofferberth,
  Lesanovsky, Fischer, Schumm, and Schmiedmayer}}]{hofferberth_1dcoherence}
\bibinfo{author}{\bibfnamefont{S.}~\bibnamefont{Hofferberth}},
  \bibinfo{author}{\bibfnamefont{I.}~\bibnamefont{Lesanovsky}},
  \bibinfo{author}{\bibfnamefont{B.}~\bibnamefont{Fischer}},
  \bibinfo{author}{\bibfnamefont{T.}~\bibnamefont{Schumm}}, \bibnamefont{and}
  \bibinfo{author}{\bibfnamefont{J.}~\bibnamefont{Schmiedmayer}},
  \bibinfo{journal}{Nature} \textbf{\bibinfo{volume}{449}},
  \bibinfo{pages}{324} (\bibinfo{year}{2007}).

\bibitem[{\citenamefont{Mazets et~al.}(2008)\citenamefont{Mazets, Schumm, and
  Schmiedmayer}}]{Mazets_breakinteg_PhysRevLett}
\bibinfo{author}{\bibfnamefont{I.~E.} \bibnamefont{Mazets}},
  \bibinfo{author}{\bibfnamefont{T.}~\bibnamefont{Schumm}}, \bibnamefont{and}
  \bibinfo{author}{\bibfnamefont{J.}~\bibnamefont{Schmiedmayer}},
  \bibinfo{journal}{Phys. Rev. Lett.} \textbf{\bibinfo{volume}{100}},
  \bibinfo{pages}{210403} (\bibinfo{year}{2008}).

\bibitem[{\citenamefont{Bongs et~al.}(2001)\citenamefont{Bongs, Burger,
  Dettmer, Hellweg, Arlt, Ertmer, and Sengstock}}]{Bongs_waveguide_PhysRevA}
\bibinfo{author}{\bibfnamefont{K.}~\bibnamefont{Bongs}},
  \bibinfo{author}{\bibfnamefont{S.}~\bibnamefont{Burger}},
  \bibinfo{author}{\bibfnamefont{S.}~\bibnamefont{Dettmer}},
  \bibinfo{author}{\bibfnamefont{D.}~\bibnamefont{Hellweg}},
  \bibinfo{author}{\bibfnamefont{J.}~\bibnamefont{Arlt}},
  \bibinfo{author}{\bibfnamefont{W.}~\bibnamefont{Ertmer}}, \bibnamefont{and}
  \bibinfo{author}{\bibfnamefont{K.}~\bibnamefont{Sengstock}},
  \bibinfo{journal}{Phys. Rev. A} \textbf{\bibinfo{volume}{63}},
  \bibinfo{pages}{031602(R)} (\bibinfo{year}{2001}).

\bibitem[{\citenamefont{Khaykovich and
  Malomed}(2006)}]{Khaykovich_quinticsol_PhysRevA}
\bibinfo{author}{\bibfnamefont{L.}~\bibnamefont{Khaykovich}} \bibnamefont{and}
  \bibinfo{author}{\bibfnamefont{B.~A.} \bibnamefont{Malomed}},
  \bibinfo{journal}{Phys. Rev. A} \textbf{\bibinfo{volume}{74}},
  \bibinfo{pages}{023607} (\bibinfo{year}{2006}).

\bibitem[{\citenamefont{Donley et~al.}(2001)\citenamefont{Donley, Claussen,
  Cornish, Roberts, Cornell, and Wieman}}]{jila:nova}
\bibinfo{author}{\bibfnamefont{E.~A.} \bibnamefont{Donley}},
  \bibinfo{author}{\bibfnamefont{N.~R.} \bibnamefont{Claussen}},
  \bibinfo{author}{\bibfnamefont{S.~L.} \bibnamefont{Cornish}},
  \bibinfo{author}{\bibfnamefont{J.~L.} \bibnamefont{Roberts}},
  \bibinfo{author}{\bibfnamefont{E.~A.} \bibnamefont{Cornell}},
  \bibnamefont{and} \bibinfo{author}{\bibfnamefont{C.~E.}
  \bibnamefont{Wieman}}, \bibinfo{journal}{Nature}
  \textbf{\bibinfo{volume}{412}}, \bibinfo{pages}{295} (\bibinfo{year}{2001}).

\bibitem[{\citenamefont{W{\"u}ster et~al.}(2005)\citenamefont{W{\"u}ster, Hope,
  and Savage}}]{wuester:nova}
\bibinfo{author}{\bibfnamefont{S.}~\bibnamefont{W{\"u}ster}},
  \bibinfo{author}{\bibfnamefont{J.~J.} \bibnamefont{Hope}}, \bibnamefont{and}
  \bibinfo{author}{\bibfnamefont{C.~M.} \bibnamefont{Savage}},
  \bibinfo{journal}{Phys. Rev. A} \textbf{\bibinfo{volume}{71}},
  \bibinfo{pages}{033604} (\bibinfo{year}{2005}).

\bibitem[{\citenamefont{W{\"u}ster et~al.}(2007)\citenamefont{W{\"u}ster,
  D{\c{a}}browska-W{\"u}ster, Bradley, Davis, Blakie, Hope, and
  Savage}}]{wuester:nova2}
\bibinfo{author}{\bibfnamefont{S.}~\bibnamefont{W{\"u}ster}},
  \bibinfo{author}{\bibfnamefont{B.~J.}
  \bibnamefont{D{\c{a}}browska-W{\"u}ster}},
  \bibinfo{author}{\bibfnamefont{A.~S.} \bibnamefont{Bradley}},
  \bibinfo{author}{\bibfnamefont{M.~J.} \bibnamefont{Davis}},
  \bibinfo{author}{\bibfnamefont{P.~B.} \bibnamefont{Blakie}},
  \bibinfo{author}{\bibfnamefont{J.~J.} \bibnamefont{Hope}}, \bibnamefont{and}
  \bibinfo{author}{\bibfnamefont{C.~M.} \bibnamefont{Savage}},
  \bibinfo{journal}{Phys. Rev. A} \textbf{\bibinfo{volume}{75}},
  \bibinfo{pages}{043611} (\bibinfo{year}{2007}).

\bibitem[{\citenamefont{Milstein et~al.}(2003)\citenamefont{Milstein, Menotti,
  and Holland}}]{holland:burst}
\bibinfo{author}{\bibfnamefont{J.~N.} \bibnamefont{Milstein}},
  \bibinfo{author}{\bibfnamefont{C.}~\bibnamefont{Menotti}}, \bibnamefont{and}
  \bibinfo{author}{\bibfnamefont{M.~J.} \bibnamefont{Holland}},
  \bibinfo{journal}{New Journal of Physics} \textbf{\bibinfo{volume}{5}},
  \bibinfo{pages}{52} (\bibinfo{year}{2003}).

\bibitem[{\citenamefont{Drummond and Brand}(2016)}]{dummond:twamctdh:comment}
\bibinfo{author}{\bibfnamefont{P.~D.} \bibnamefont{Drummond}} \bibnamefont{and}
  \bibinfo{author}{\bibfnamefont{J.}~\bibnamefont{Brand}}
  (\bibinfo{year}{2016}), \eprint{arXiv:1610.07633v1}.

\bibitem[{\citenamefont{Sakmann and Kasevich}(2017)}]{sakman:twamctdh:comment}
\bibinfo{author}{\bibfnamefont{K.}~\bibnamefont{Sakmann}} \bibnamefont{and}
  \bibinfo{author}{\bibfnamefont{M.}~\bibnamefont{Kasevich}}
  (\bibinfo{year}{2017}), \eprint{arXiv:1702.01211v2}.

\bibitem[{\citenamefont{Cosme et~al.}(2016)\citenamefont{Cosme, Weiss, and
  Brand}}]{Cosme_com_motion}
\bibinfo{author}{\bibfnamefont{J.~G.} \bibnamefont{Cosme}},
  \bibinfo{author}{\bibfnamefont{C.}~\bibnamefont{Weiss}}, \bibnamefont{and}
  \bibinfo{author}{\bibfnamefont{J.}~\bibnamefont{Brand}},
  \bibinfo{journal}{Phys. Rev. A} \textbf{\bibinfo{volume}{94}},
  \bibinfo{pages}{043603} (\bibinfo{year}{2016}).

\bibitem[{\citenamefont{Weiss et~al.}(2015)\citenamefont{Weiss, Gardiner, and
  Breuer}}]{weiss_CMdiffusion}
\bibinfo{author}{\bibfnamefont{C.}~\bibnamefont{Weiss}},
  \bibinfo{author}{\bibfnamefont{S.~A.} \bibnamefont{Gardiner}},
  \bibnamefont{and} \bibinfo{author}{\bibfnamefont{H.-P.}
  \bibnamefont{Breuer}}, \bibinfo{journal}{Phys. Rev. A}
  \textbf{\bibinfo{volume}{91}}, \bibinfo{pages}{063616}
  (\bibinfo{year}{2015}).




\end{thebibliography}
\end{document}